# *Ab initio* Understanding of the Pseudogap in Cuprate High Temperature Superconductors via the Fluctuating Bond Model

R. A. Nistor[1,2]

[1] *Dep. Applied Mathematics, University of Western Ontario, ON N6A 5B7 Canada*

G. J. Martyna[2], D. M. Newns[2], C. C. Tsuei[2]

[2] *IBM T.J. Watson Research Center, Yorktown Heights, NY 10598 USA*

M. H. Müser[1,2,3]

[3] *Universität des Saarlandes, Materialgerechtes Design und Werkstoffinformatik, Campus, Geb. C6.3, 66123 Saarbrücken, Germany*

## Abstract

Understanding the origin of the pseudogap is an essential step towards elucidating the pairing mechanism in the cuprate superconductors. Recently there has been strong experimental evidence showing that C4 symmetry breaking occurs on formation of the pseudogap. This form of symmetry-breaking was predicted by the Fluctuating Bond Model (FBM), an empirical model based on a strong, local coupling of electrons to the square of the planar oxygen vibrator amplitudes. In this paper we approach the FBM theory from a new direction, starting from *ab initio* molecular dynamics simulations. The simulations demonstrate a doping-dependent instability of the in-plane oxygens towards displacement off the Cu-O-Cu bond axis. From these results and perturbation theory we derive an improved and quantitative form of the Fluctuating Bond Model. A mean field solution of the FBM leads to C4 symmetry breaking in the oxygen vibrational amplitudes, and to a *d*-type pseudogap in the electronic spectrum, the features linked by recent experimental data. The phase diagram of the pseudogap derived from mean field theory, its doping- and temperature-dependences, including the phase boundary $T^*$, agree well with experimental data. We extend the theory to include the long range Coulomb interaction on the same basis as the FBM interaction. When the long-range Coulomb interaction is included in the FBM, a CDW instability in the charge channel is predicted which explains the nanoscale, rather than spatially uniform, behavior of the C4 symmetry-breaking. Taking the CDW into account, with the theoretical $k$-dependence of the pseudogap, enables the Fermi Surface arc phenomenon to be understood.

After years of intensive theoretical and experimental effort, there is still no consensus as to the pairing mechanism in cuprate high temperature superconductors (HTS), nor on the origin of the pseudogap (PG) [1], which needs to be an integral part of the eventual solution to the HTS problem. Perhaps it is time to extend our thinking beyond some of the most attractively simple models explored over the last 20 years, such as for example the large-$U$ Hubbard model [2] or models with linear electron-phonon coupling [3].

Symmetry-breaking permeates all branches of physics, and its study often allows us to gain insight into the nature of the underlying physical phenomena. This approach can be invoked in order to throw light on the origin of the pseudogap in cuprate high temperature superconductors (HTS), and ultimately to help elucidate the nature of the pairing mechanism in these materials. Recent evidence shows that the pseudogap is associated with the presence of C4 symmetry breaking [4, 5], i.e. the $a$ and $b$ directions in the $CuO_2$ plane become nonequivalent. The nature of the C4 symmetry breaking at low temperature is revealed by atomic resolution STM studies [6], which show that it is associated with the oxygens in the $CuO_2$ plane, the oxygens in the $x$-directed Cu-O-Cu bonds differing from the oxygens in $y$-directed Cu-O-Cu bonds both in their electronic properties and in their vibrational amplitudes. At low temperature and under conditions where dopant nonuniformity creates electronic nanoscale inhomogeneity [7], C4 symmetry breaking has been found to be coterminous in space with the regions where the PG is present [5]. At high temperatures, around the temperature $T^*$ at which the PG appears, C4 symmetry breaking has been found experimentally [4] to turn on at the temperature $T^*$, and the simplest assumption is that this high temperature C4 symmetry breaking has the same origin as that revealed by the low temperatiure STM work. The most straightforward reading of the evidence is then that the PG is a symmetry breaking phenomenon, in which the oxygens in $x$-directed Cu-O-Cu bonds become electronically and vibrationally different from the oxygens in $y$-directed Cu-O-Cu bonds.

In an earlier study [8] which introduced an empirical model for cuprate HTS, the Fluctuating Bond Model, C4 symmetry breaking was predicted prior to its initial observation [7], and indeed in this theory oxygens in $x$- and $y$- directed Cu-O-Cu bonds were shown to become electronically and vibrationally distinct. The basis of the FBM is a *nonlinear* coupling between the the vibrational coordinates of the oxygen atoms in the $CuO_2$ plane and the electron system, distinguishing it from the *linear* electron-lattice coupling [9] in conventional

BCS superconductivity. The FBM naturally explained [8] the $d$-wave superconducting properties of HTS [10], where earlier related approaches [11, 12] were unsatisfactory as models of HTS as they predicted $s$-wave. Reference [13] explores superconductivity arising from even higher order couplings. For the FBM to be convincing as a mechanism for C4 symmetry breaking, it needs to be shown that electron-lattice coupling, and in particular nonlinear coupling, is significant in HTS, and that the FBM can describe the experimental behavior of the C4 symmetry breaking/PG phenomenon. These are the objectives of this paper.

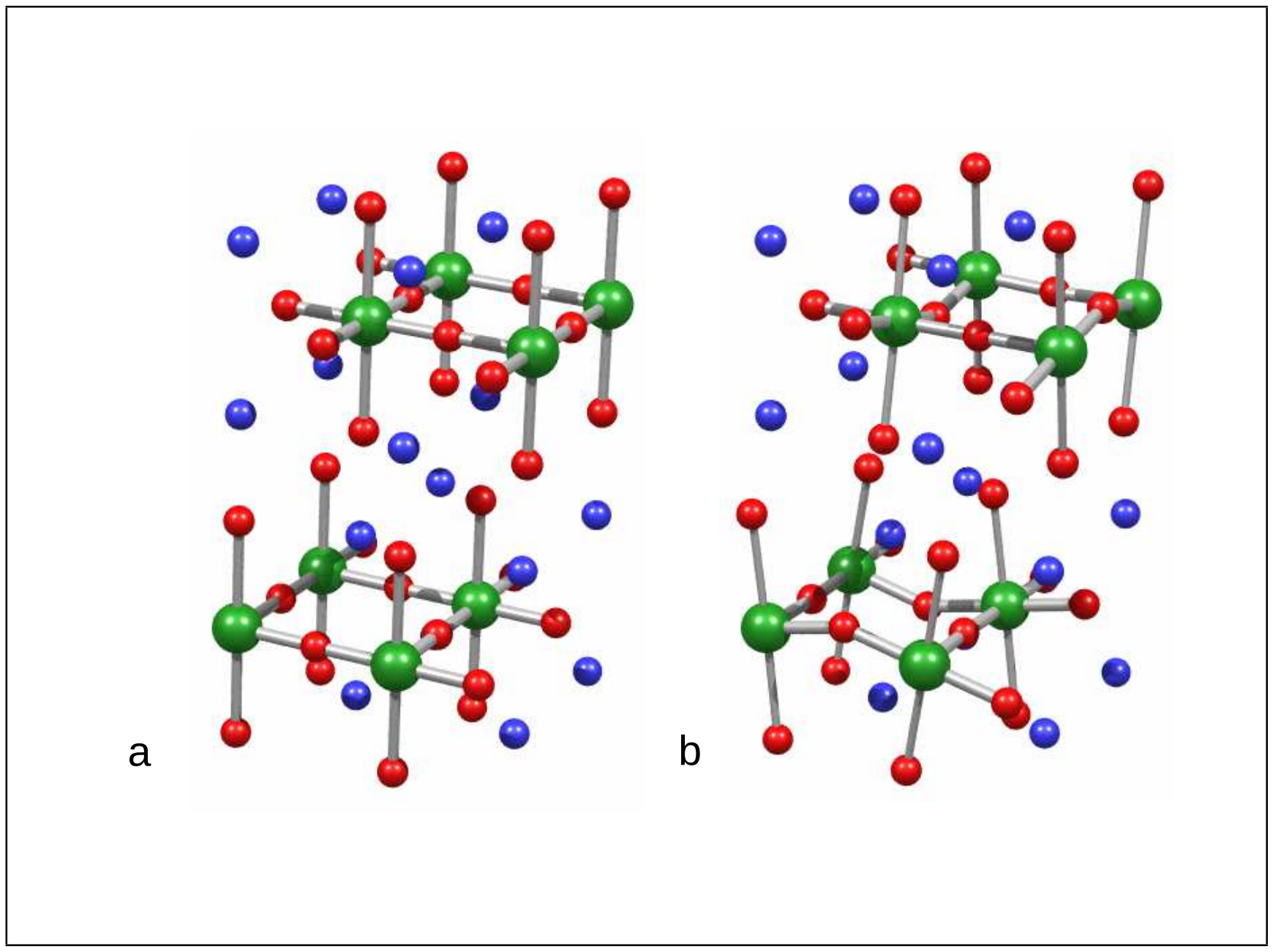

FIG. 1: *ab initio* Molecular Dynamics calculation at $T = 4\,\mathrm{K}$ of the structure of metallic $La_2CuO_4$ (214) [21], blue spheres, La, green spheres, Cu, red spheres, O. **a**, Undistorted setup structure, **b**, Equilibrated structure showing vertical displacements of the planar oxygens corresponding to rotations of $CuO_6$ octahedra about alternate $x$- and $y$- axes in planes stacked along the $c$-axis - the LTT structure found at low temperature in metallic 214 phases.

The significance of electron-lattice coupling in cuprate HTS can be inferred from experimental evidence regarding the pairing mechanism [3], such as the universal doping-dependent oxygen isotope shift [14, 15] and the superconductivity-induced softening in oxygen vibration frequency [16, 17]. However, conventional linear electron-phonon coupling is not straightforwardly related to C4 symmetry breaking and produces a *doping-independent* isotope shift. Also *ab initio* calculations do not find a strong conventional *linear* electron-lattice coupling in the cuprates [18, 19]. We shall see that these difficulties can be resolved if we extend our thinking to include *nonlinear* electron-lattice coupling.

Fundamental grounds for emphasizing nonlinear electron-lattice coupling emerge when O motion transverse to the axis of the Cu-O-Cu bond in the $CuO_2$ plane (the transverse modes turn out to be the key ones) is considered. We argue from the symmetric environment of the Cu-O-Cu bond in cuprates that the local effect of the transverse O motion on the electrons should be independent of the O displacement's sign, and hence *second order* in the O displacement. Linear coupling of the bond-transverse O modes should only come in as a relatively long range piece depending on nonuniversal structural elements which violate inversion symmetry in the bond axis - providing the basis for small linear coupling as determined by *ab initio* methods [18, 19].

We shall show in this paper (a) using *ab initio* molecular dynamics that there is a strong nonlinear coupling between electrons and the bond-transverse O vibrations, which can be formally expressed in terms of a quantitatively parameterized FBM, and (b) that this coupling leads to a natural explanation of C4 symmetry breaking and of the associated pseudogap and its phenomenology. Finally we show that including the long-range Coulomb interaction (LRCI) in the FBM explains the charge density wave (CDW) which modulates the symmetry-breaking.

## I. *AB INITIO* FOUNDATION OF THE FBM

The powerful technique of *ab initio* molecular dynamics (AIMD) [20] solves the ionic equations of motion on a first principles Born-Oppenheimer potential energy (PE) surface, a conceptual step forward from the empirical PE surface used in conventional MD. The *ab initio* PE surface is obtained by solution of the many-electron Schrodinger equation in local density approximation, most often augmented by gradient correction. This technique is well suited to the present problem because it avoids the constraint of a linearized electron-lattice interaction. We start by using AIMD to show that the oxygens in the $CuO_2$ plane are unstable, leading to the observed symmetry-breaking, then identify the cause of the instability, which leads naturally to formulating the FBM.

First we consider the the oxygen instability leading to the well-established Low Temperature Tetragonal (LTT) structure [21] found in underdoped metallic 214 materials, which was early on linked to non-linear electron lattice coupling [22]. Figure 1 shows that the LTT structure is indeed predicted by AIMD for metallic $La_2CuO_4$ at $T = 4\,K$. In a $CuO_2$ plane the oxygens in say the $x$-directed bonds are displaced, half up and half down, along the $z$-axis, while in the next $CuO_2$ plane the oxygens in the $y$-directed bonds are displaced, etc.. Hence each $CuO_2$ plane breaks C4 symmetry, but the alternation of bond distortion between $x$- and $y$- directed bonds ensures overall tetragonal symmetry.

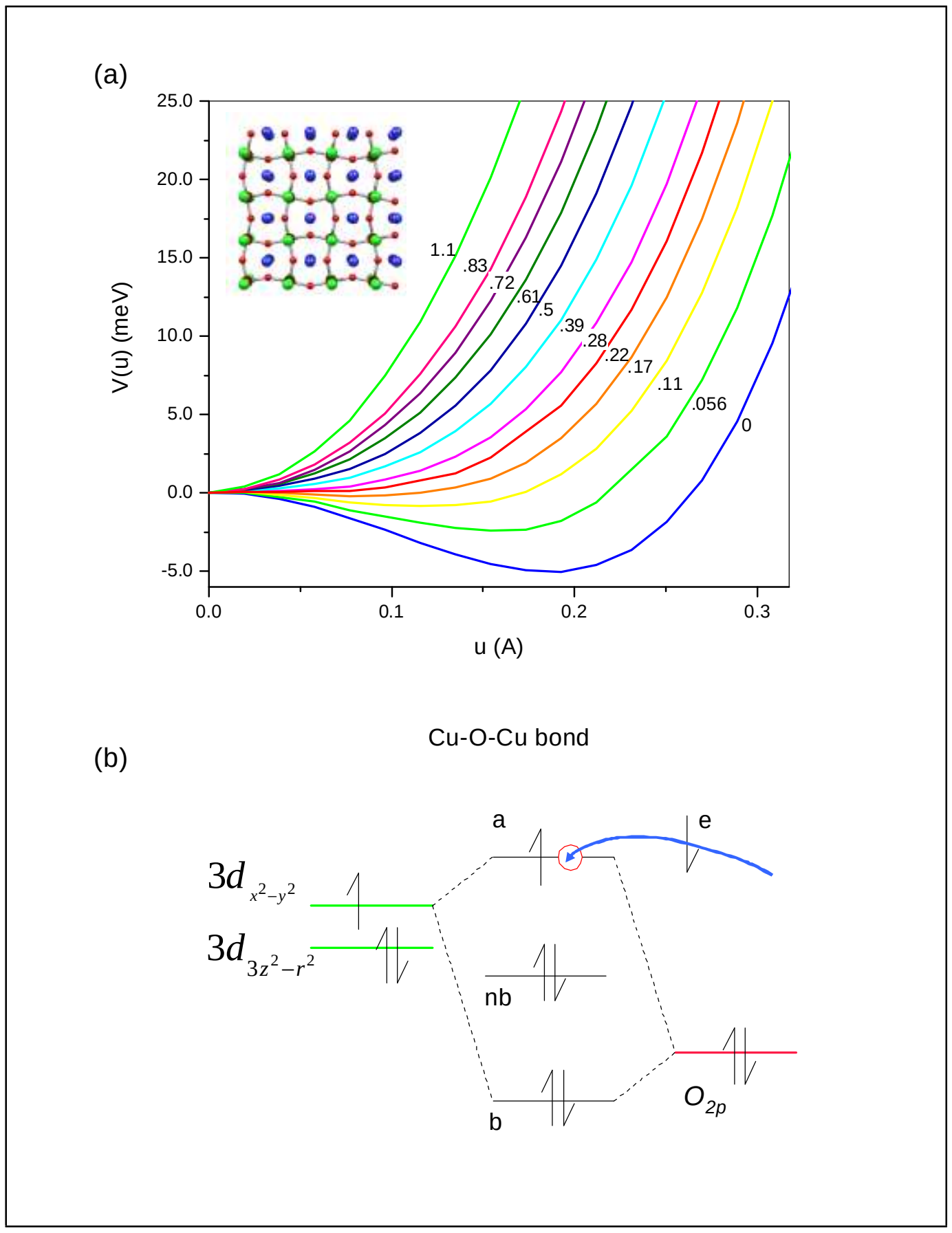

FIG. 2: (a) PE curves $V(u)$ for oxychloride material as a function of oxygen in-plane distortion $u$ (see inset) for different dopings (see labelling on curves). Inset color code, blue spheres, Ca/Na, green, Cu, red, O. Doping is implemented by fractional substitution of Na for Ca. (b) Breaking of the Cu-O-Cu bond by electron addition. Since bonding "b"and nonbonding "nb" levels are occupied, bond strength depends on holes in partially-occupied antibonding "a" level. Adding an electron to the "a" level will eliminate bond strength leading to off-axis PE surface minimum for oxygen atom.

To analyze the mechanism of the oxygen instability we turn to the oxychloride system $Ca_{2-x}Na_xCuO_2Cl_2$, which is computationally advantageous and whose doping can be controlled via the Na fraction $x$. In this system we have calculated the oxygen potential energy (PE) surfaces as a function of doping (see Fig. 2a). In the oxychloride the AIMD calculation (see inset Fig. 2a) shows that in contrast to the 214 material the oxygen instability is in the $xy$-plane, in agreement with experiment [6] (though AIMD cannot capture the experiment's non-Born-Oppenheimer features).

What is remarkable in Fig. 2a, supporting an electronic origin for the oxygen instability, is that the PE surfaces are strongly dependent on doping. At low doping the PE minimum is off the bond axis (leading to the Fig. 2a inset distortion), transitioning at high doping to an on-axis minimum (stability of oxygen on the bond axis). The AIMD results can be parameterized in the form

$$V(u) = (\chi + Vp)\,\frac{u^2}{2} + \frac{w}{8}u^4, \tag{1}$$

where $u$ is oxygen displacement from the bond axis. In (1) the force constant $\chi$ is negative when doping $p$ is zero (oxygen unstable at zero doping), while the electron-lattice coupling $V$ is positive, representing stabilization of the intrinsically unstable Cu-O-Cu bond with increasing hole doping. The positive quartic term $w$ confines the oxygen atom in the local lattice cage for the unstable cases. Eq. (1) embodies the result that the electron-lattice coupling in Fig. 2a goes as the *square* of the oxygen displacement, as was argued above.

It is very helpful to interpret the *ab initio* results in Fig. 2a and Eq. (1) in terms of a local chemical bonding energy level picture. In a two-atom bond such as that in $H_2$ there is a low energy bonding orbital, which is doubly occupied, and a higher energy antibonding orbital, which is empty. The strength of the bond is optimum with these occupations; the bond strength would be zero if the occupations were zero, and also if both bonding and antibonding levels were both occupied. In Fig. 2b, we sketch the local chemical energy level picture for the three-atom Cu-O-Cu $\sigma$-bond. Because there are three atoms, there are now bonding, non-bonding, and antibonding levels. Again if all levels are filled, the bond strength is zero. The bonding and nonbonding levels are filled, so the bond strength relies entirely on partial filling of the the antibonding level, which in an undoped system involves only 1/2 hole per bond, i.e. the antibonding level is 3/4 filled. This is a very weak bond and is in fact unstable, as seen in Fig. 2a, where the energy minimum is off-axis for zero doping. As the hole number in the antibonding orbital is increased by doping, the bond will become stable, exactly as seen in the AIMD results in Fig 2a, where the energy minimum moves to the bond axis. Direct *ab initio* support for this local chemical bonding picture has in fact been obtained in a set of calculations on linear molecules of the type X-Cu-O-Cu-X [23]. In these calculations the Cu-O-Cu bond is "doped" by the choice of electron withdrawing/electron donating group X. Doping with holes/electrons stabilizes/destabilizes the Cu-O-Cu bond just as shown in Fig. 2a for a cuprate system. The added electrons are found to go into the antibonding orbital, just as sketched in the chemical picture of Fig 2b.

## II. DERIVATION OF THE FBM HAMILTONIAN

As we have discussed above, the Fig. 2 AIMD results support a coupling between oxygen vibrator force constant and electron occupation of the antibonding orbital. We now give a more formal derivation of this form of coupling (see Appendix A). The approach requires a model. We start from the 3-band Emery tight-binding model based on Cu $3d_{x^2-y^2}$ and the oxygen $2p_x/2p_y$ orbitals that have $\sigma$ symmetry in the Cu-O-Cu bond (see Fig. 3). The key parameters in the 3-band model are the $pd$ hopping matrix elements $t_{pd}$, and the $p$ to $d$ energy gap $\epsilon_{pd} > 0$. In this paper we work with the more tractable and widely used 1-band model [24] rather than the 3-band model. The basis set in the 1-band model consists of a single $3d_{x^2-y^2}$ orbital per Cu atom located at site $i$ on the square Cu Bravais lattice in the $CuO_2$ plane.

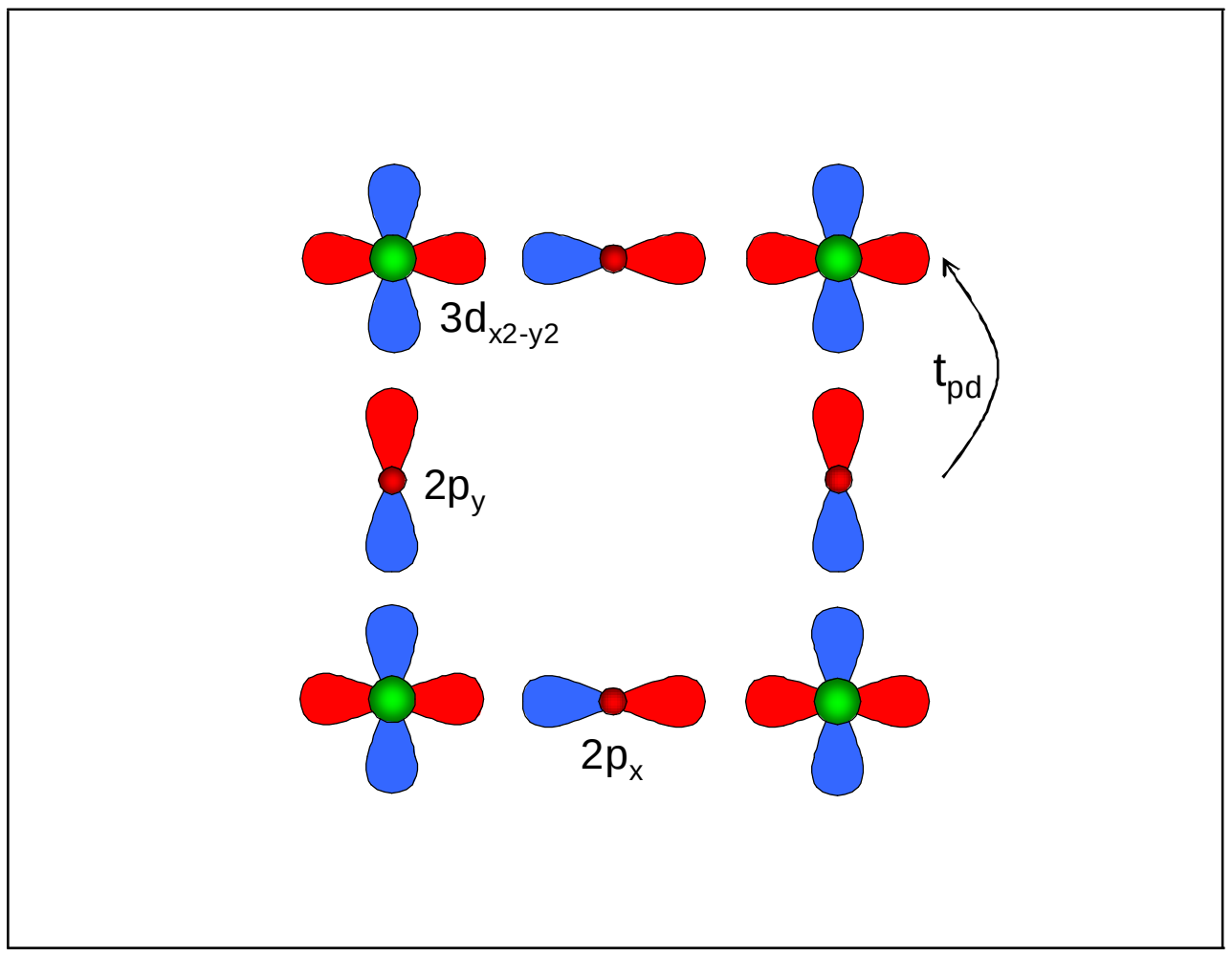

FIG. 3: $2p_x$, $2p_y$ and $3d_{x^2-y^2}$ orbitals in $CuO_2$ plane, illustrating $2p$ to $3d$ hopping integral $t_{pd}$.

A projection procedure (Appendix A) enables approximate passage from the 3-band to the 1-band model, which becomes

$$\widetilde{H}^d = \sum_{i,\sigma} \epsilon_d n_{i\sigma} - \sum_{\langle i,j\rangle,\sigma} \frac{t^2_{p_{ij}d}}{\epsilon_{p_{ij}d}} \left(c^+_{i,\sigma} c_{j,\sigma} + c^+_{j,\sigma} c_{i,\sigma}\right) + \sum_{\langle i,j\rangle,\sigma} \frac{t^2_{p_{ij}d}}{\epsilon_{p_{ij}d}} \left(n_{i\sigma} + n_{j\sigma}\right), \tag{2}$$

(here $c^+_{i,\sigma}$ ($c_{i,\sigma}$) is the creation (destruction) operator for the $3d_{x^2-y^2}$ orbital of spin $\sigma$ on site $i$, with number operators $n_{i,\sigma} = c^+_{i,\sigma} c_{i,\sigma}$, and $\epsilon_d$ is the $3d_{x^2-y^2}$ orbital energy). A sum over $\langle ij \rangle$ implies that each nearest-neighbor bond $ij$ appears only once in the sum. In (2) the 3-band model parameters $t_{pd}$, and $\epsilon_{pd}$ have been made bond-dependent.

The key physical content is seen in the second term of (2). This term describes a super-exchange hopping $t$ between nearest-neighbor Cu atoms $i$ and $j$ driven by electrons hopping from $i$ to the intermediate $p$-orbital via $t_{pd}$ and then from the intermediate $p$-orbital to $j$ via another $t_{pd}$ matrix element (and the reverse). There is also an energy shift in the $3d_{x^2-y^2}$ orbitals (third term in (2)) due to the process where after reaching the intermediate $p$-orbital from $i$ an electron hops back again to $i$ instead of going on to $j$.

A vibrational displacement $u$ of the oxygen, typically transverse to the Cu-O-Cu bond axis, will modify the $pd$ hopping matrix elements $t_{pd}$. The modification will tend to reduce the $pd$ overlap, hence will be of the form (taking $t_{pd} > 0$, as in Fig. 3)

$$t_{pd} \rightarrow t_{pd} - v_{pd} u^2, \quad \text{where } v_{pd} > 0. \tag{3}$$

Inserting this approximation into (2), and expanding only as far as the second order in $u$ we obtain the nonlinear electron-vibrator coupling model

$$\begin{aligned} \widetilde{H}^d = {} & \epsilon'_d \sum_{i,\sigma} n_{i\sigma} - t \sum_{\langle i,j\rangle,\sigma} \left(c^+_{i,\sigma} c_{j,\sigma} + c^+_{j,\sigma} c_{i,\sigma}\right) \\ & - \frac{v}{2\sqrt{2}} \sum_{\langle i,j\rangle,\sigma} \left((n_{i\sigma} + n_{j\sigma}) - \left(c^+_{i,\sigma} c_{j,\sigma} + c^+_{j,\sigma} c_{i,\sigma}\right)\right) u^2_{ij}, \end{aligned} \tag{4}$$

where $\epsilon'_d = \epsilon_d + 2t$ (the trivial shift $2t$ in the $d$-orbital energy will subsequently be ignored), $t = t^2_{pd}/\epsilon_{pd}$ is the 1-band tight binding hopping matrix element, and the electron-vibrator coupling matrix element $v$ is defined by $v/2\sqrt{2} = 2t_{pd}v_{pd}/\epsilon_{pd}$ [8] (in this paper we define the vibrator and electron spin degeneracies as 1 and 2 respectively). The coupling $v$ is seen to be **positive.** Our original empirical model [8] contained only the hopping terms in the coupling, and missed the number operator terms (we shall see that this modification has little effect on a *uniform* pseudogap). Note that a similar electron-vibrator coupling term occurs if the coupling originates from $\epsilon_{pd}$ (see Appendix A), instead of from $t_{pd}$ (as in Eq. (3)).

A neat way to express the nonlinear electron-vibrator coupling term is to introduce the antibonding orbital $[c_{i,\sigma} - c_{j,\sigma}]/\sqrt{2}$ for the $ij$ bond. The number operator for this bond, summed over spin, is defined as $Q_{ij}$

$$\begin{aligned} Q_{ij} &= \sum_{\sigma} \left( \frac{[c_{i,\sigma} - c_{j,\sigma}]}{\sqrt{2}} \right)^{+} \left( \frac{[c_{i,\sigma} - c_{j,\sigma}]}{\sqrt{2}} \right) \\ &= \frac{1}{2} \sum_{\sigma} \left( (n_{i\sigma} + n_{j\sigma}) - \left( c^{+}_{i,\sigma} c_{j,\sigma} + c^{+}_{j,\sigma} c_{i,\sigma} \right) \right) , \end{aligned} \tag{5}$$

and is seen to be the electronic factor in the coupling term in (4), which can now be written compactly as

$$\widetilde{H}^{d} = \epsilon'_{d} \sum_{i,\sigma} n_{i\sigma} - t \sum_{\langle i,j \rangle,\sigma} \left( c^{+}_{i,\sigma} c_{j,\sigma} + c^{+}_{j,\sigma} c_{i,\sigma} \right) - \frac{v}{\sqrt{2}} \sum_{\langle i,j \rangle} Q_{ij} u^{2}_{ij}. \tag{6}$$

Some further additions are required in order to arrive at a complete and realistic FBM Hamiltonian (see Appendix B). The nonlinear coupling in (6) can in principle have the effect of *deconfining* the vibrating oxygen (giving it a parabolic convex-downwards PE surface). The *ab initio* PE curves in Fig. 2a and Eq. (1) show that the oxygen vibrator is in fact confined by a quartic-type potential, which therefore needs to be included. We also need to add the standard kinetic energy and parabolic potential energy terms for the vibrator. The electronic Hamiltonian also needs refining by including next nearest hopping terms $t'$ and next-next nearest neighbor terms $t''$ ($t'$ is especially important as it sets the doping where the van Hove singularity peak in the DOS is located at the Fermi level).

The resulting FBM Hamiltonian Eq. (7) (see Appendix B) is written in mixed representation electronically

$$H = \sum_{\mathbf{k},\sigma} \epsilon_{\mathbf{k}} n_{\mathbf{k},\sigma} + \sum_{\langle i,j \rangle} \left[ \frac{p^{2}_{ij}}{2M} + \frac{\chi_{0}}{2} u^{2}_{ij} \right] + \frac{w}{8} \sum_{\langle i,j \rangle} u^{4}_{ij} - \frac{v}{\sqrt{2}} \sum_{\langle i,j \rangle} Q_{ij} u^{2}_{ij}. \tag{7}$$

In Equation (7) $c^{+}_{\mathbf{k},\sigma}$ ($c_{\mathbf{k},\sigma}$) are the creation (destruction) operators for the band states of wavevector $\mathbf{k}$, obtained by diagonalizing the tight binding model with the hopping matrix elements $t, t'$, and $t''$, and $n_{\mathbf{k},\sigma} = c^{+}_{\mathbf{k},\sigma} c_{\mathbf{k},\sigma}$ is the corresponding number operator. The variables $p_{ij}$, $u_{ij}$ are the conjugate momentum and position coordinates of oxygen in bond $ij$, $M$ is oxygen mass and $\chi_0$ the bare oxygen force constant. An Einstein model is assumed, so inter-vibrator interactions are ignored. $w$ is the quartic interaction and $v$ the (positive) electron-vibrator coupling constant. There is a characteristic coupling energy $K = v^2/w$ in the model, related to the pairing energy [8].

Table I: FBM Parameters for 214 and Oxychloride Materials

| Parameter | vib. $xy \perp$ to bond | vib. $z \perp$ to bond |
|---|---|---|
| $v_{214}$ (au) | 0.016 | 0.017 |
| $v_{\mathrm{oxy}}$ (au) | 0.018 | 0.020 |
| $w_{214}$ (au) | 0.053 | 0.122 |
| $w_{\mathrm{oxy}}$ (au) | 0.090 | 0.106 |
| $8K_{214}$ (eV) | 1.12 | 0.54 |
| $8K_{\mathrm{oxy}}$ (eV) | 0.80 | 0.83 |

The coupling term (last term) in Eq.(7) can be interpreted in terms of the Fig. 2b chemical picture, it states that if we increase the occupation $Q_{ij}$ of the antibonding orbital in bond $ij$ (see Eq. (5)) then the Cu-O-Cu bond $ij$ is softened. Looking at the hopping terms within $Q_{ij}$ in (5), then it is seen that increasing the vibrational amplitude in bond $ij$ changes the nearest-neighbor hopping term so as to reduce the effective hopping $|t|$ (see (6)) in bond $ij$.

The remaining terms in Eq.(7) are as follows. The first term is the electronic band energy. The second term represents the harmonic part of the oxygen vibrational Hamiltonian, to which is added the third term, a quartic interaction needed to confine the oxygen and derived from the PE curves in Fig 2a. The Hamiltonian Eq.(7), termed FBMII, differs from the original FBM model [8] (now termed FBMI) in the presence of the number operator terms in $Q_{ij}$ (Eq. (5)), and, as we shall now see, in having the key parameters determined fom *ab initio* calculations.

The FBM nonlinear electron-vibrator coupling is especially effective at the high-density of states saddle points at X$= (\pi, 0)$ and Y$= (0, \pi)$ in the band structure. The energies at X and Y are normally degenerate, but the degeneracy is split if the vibrational amplitudes $u$ for the oxygens in $x$-directed bonds are not the same as the amplitudes in $y$-directed bonds (see Fig. 4a). This splitting can be used to determine the bare electron-lattice coupling constant $v$ in Eq. (7) by displacing the $x$-oxygens and calculating the shift in the band structure eigenvalue at X. Any effect of a global chemical potential shift due to displacing the $x$-oxygens can be removed by displacing the $y$-oxygens and subtracting the $y$-induced shift at X from the $x$-induced shift. In Appendix C we discuss in more detail how the shift in band structure energy eigenvalues $\epsilon_{\mathbf{k}}$ at the saddle points X and Y as a result of displacing the oxygens can be used to determine the coupling constant $v$. The results are collected in Table I for the 214 and oxychloride materials. The values of the quartic interaction $w$ are obtained from the quartic coefficient of the fit to the Fig. 2a curves, and similar ones for the 214 material. It is found that the coupling $v$ is relatively small for vibrational polarization along the Cu-O-Cu bond, so we only considered polarizations transverse to the bond in Table I. Repeating this calculation with the $U$-facility in the Quantum Espresso code enabled did not significantly change the results, a finding which suggests that the FBM couplings are not an artefact of neglecting electron correlation effects.

The lower section Table I shows the coupling strength $K = v^2/w$ in the FBM, which will be discussed further below. We now turn to the pseudogap results obtainable from the FBM at the mean field level.

## III. C4 SYMMETRY BREAKING AND THE PSEUDOGAP

A mean field approximation to a new Hamiltonian is often found to yield valuable insights. The nonlinear form of coupling in the FBM Eq.(7) lends itself to an unusual form of mean field theory where $u_{ij}^2$ can be replaced by its expectation value $\langle u_{ij}^2 \rangle$. The details of this mean field theory are supplied in Appendix D, and the essentials are described as follows. When $u_{ij}^2$ is replaced by its expectation value $\langle u_{ij}^2 \rangle$ the coupling term becomes $v \langle u_{ij}^2 \rangle \left( c_{i,\sigma}^+ c_{j,\sigma} + c_{j,\sigma}^+ c_{i,\sigma} \right) /2\sqrt{2}$, locally modifying the nearest neighbor hopping $t$, which appears in the one-electron Hamiltonian $-t \left( c_{i,\sigma}^+ c_{j,\sigma} + c_{j,\sigma}^+ c_{i,\sigma} \right)$, to $t \rightarrow \left( t - v \langle u_{ij}^2 \rangle /2\sqrt{2} \right)$, decreasing the hopping strength since $v > 0$.

Moreover, if the vibrational amplitude $\langle u_{ij}^2 \rangle$ were to differ between $x$- and $y$-directed bonds the nearest neighbor hopping $t$ would also differ, becoming say $t_x$ and $t_y$ respectively. Now the saddle point energies $\epsilon_X$ and $\epsilon_Y$ are given by

$$\begin{aligned} \epsilon_X &= -2(-t_x + t_y) + 4t' - 4t'', \\ \epsilon_Y &= -2(t_x - t_y) + 4t' - 4t'', \end{aligned} \qquad (8)$$

so the energies of the saddle points are split by $\epsilon_X - \epsilon_Y = 4(t_x - t_y)$. Splitting the saddle points splits the van Hove singularity [25, 26] in the density of states (DOS) (see Fig. 4a) which leads to a Peierls-like mechanism for creating the vibrational amplitude asymmetry in $\langle u_{ij}^2 \rangle$ self-consistently. This is the underlying process which leads to C4 symmetry-breaking and to the pseudogap in the FBM.

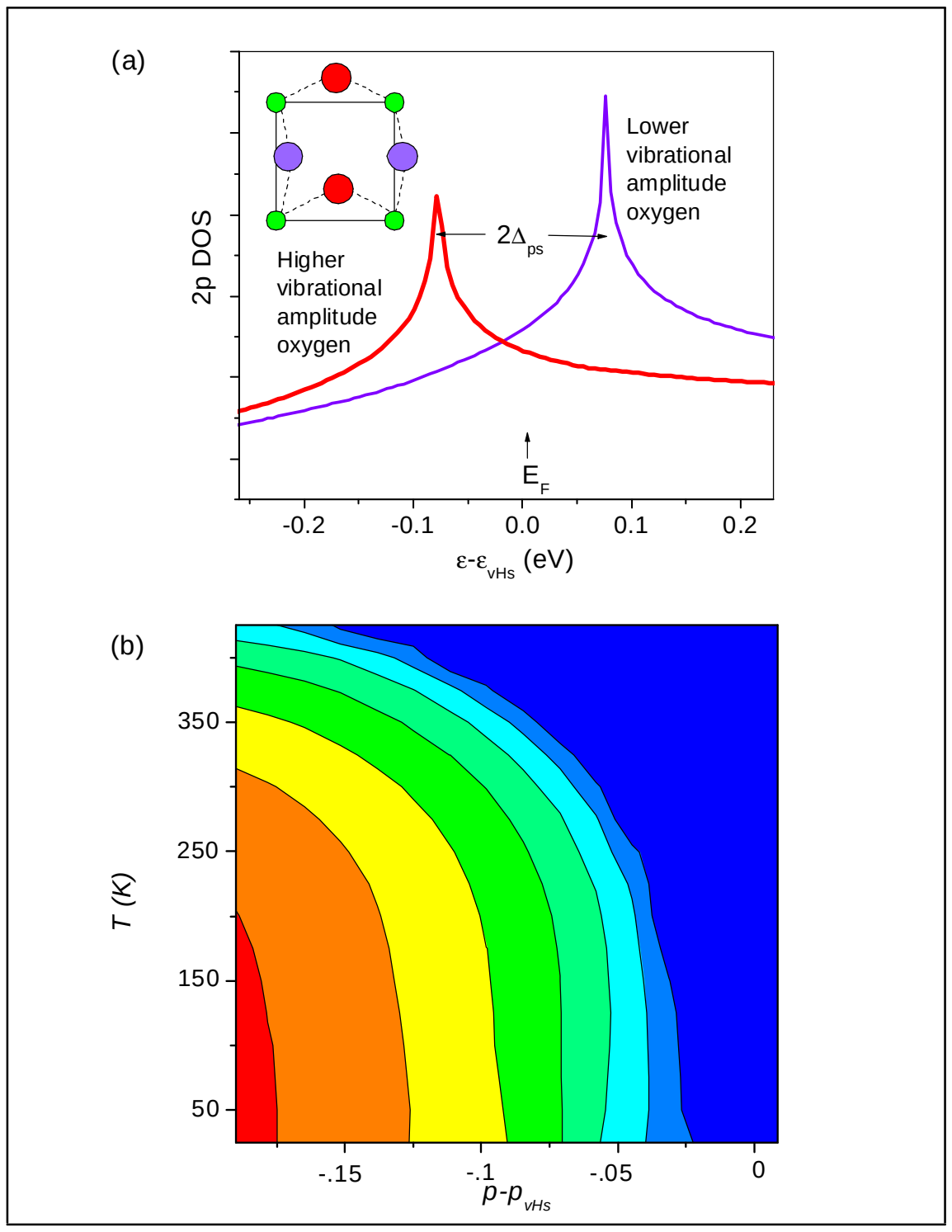

FIG. 4: a) Oxygen-projected 2p-DOS for oxygens in x-oriented (red) and y-oriented (violet) bonds.The vHs peak above the Fermi level (violet) is for the lower vibrational amplitude oxygen, and the peak below the Fermi level (red) is for the higher vibrational amplitude oxygen. For details, see Appendix A (b) Contour map of pseudogap in temperature/doping plane showing decrease with doping, and with temperature, until it vanishes at phase boundary $T^*$. Contours labelled by pseudogap $\Delta_{ps}$ in intervals of 13.75 meV. For experimental $\Delta_{ps}$ magnitudes see Ref. [30].

Applying mean field to the model Eq.(7) (Appendix D), we then expand the coupling term into two possible decouplings (a) $Q_{ij}u_{ij}^2 \rightarrow Q_{ij} \langle u_{ij}^2 \rangle$, with consequences just discussed, and (b) $Q_{ij}u_{ij}^2 \rightarrow \langle Q_{ij} \rangle u_{ij}^2$ , leading to a softening of the vibrator proportional to the number of antibonding electrons $\langle Q_{ij} \rangle$ in the bond (the Fig. 2a, 2b effect). Here for the moment we assume for simplicity that the mean field solution is translationally invariant, when the decoupling breaks the problem into two exactly soluble pieces. The (a) decoupling leads to a band structure problem which is solved to give the expectation value $\langle Q_{ij} \rangle$, which can be fed into (b) to give the softened vibrator frequency. The solution to the anharmonic oscillator problem posed by Eq.(7) with the decoupling (b) is done by expanding in a harmonic oscillator basis, leading to a value of $\langle u_{ij}^2 \rangle$ to be fed back into (a). The mean field quantities $\langle Q_{ij} \rangle$ and $\langle u_{ij}^2 \rangle$, are solved for self-consistently, further details are provided in Appendix D.

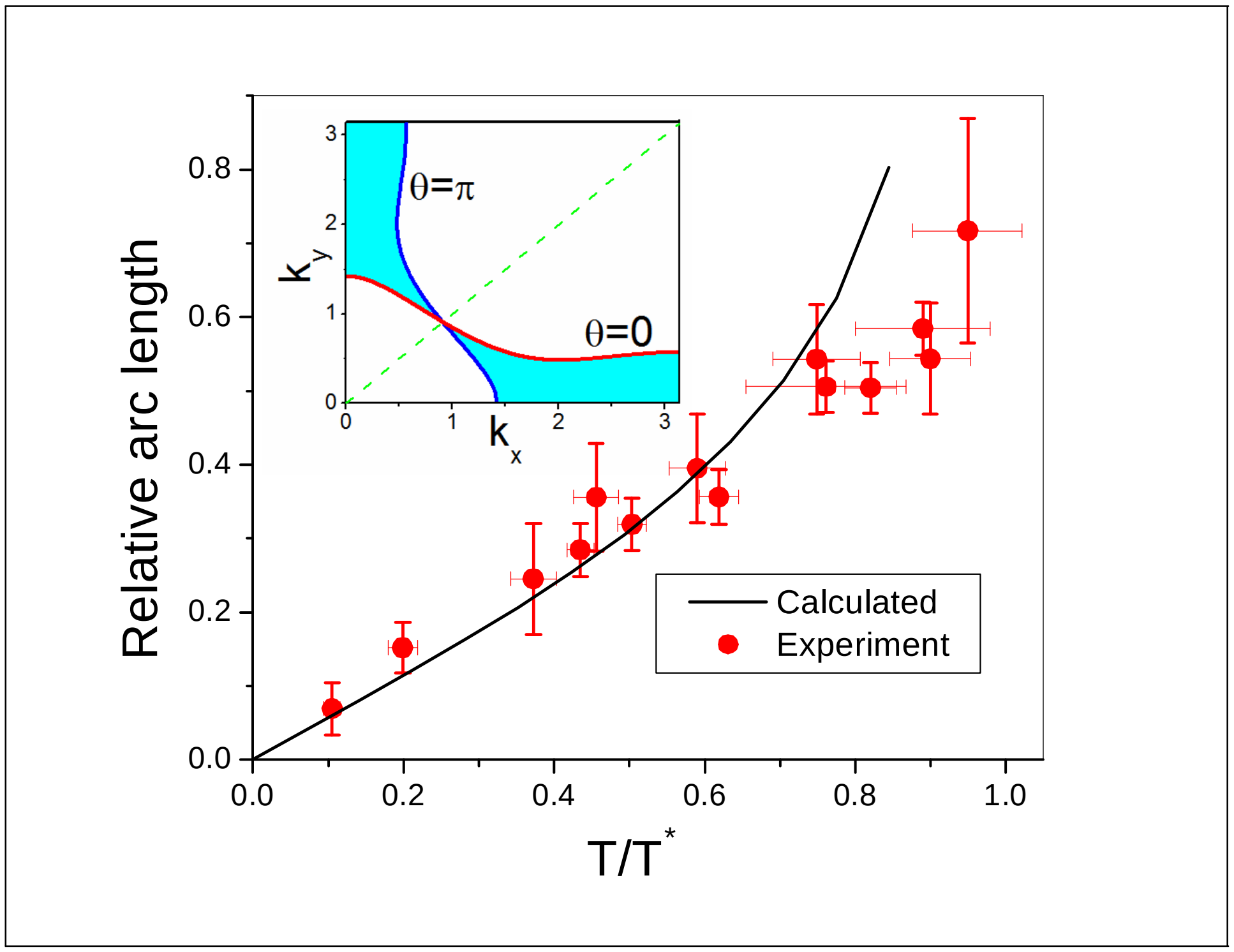

FIG. 5: Inset: two FS's for opposite sign of C4-splitting phase ($\Delta_{ps} = 74$ meV). $\theta = 0$ and $\theta = \pi$ are defined in the context of the order parameter $\Delta_{ps}(\mathbf{k}) \sim \Delta_{ps} \cos\theta \left(\cos k_x - \cos k_y\right)/2$ (see text). Colored area indicates approximate loss of definition of FS due to CDW. Main panel: Plot of FS arc length vs. temperature compared with experiment (see text).

At high temperatures the lowest free energy solution to the mean field equations preserves C4 symmetry. But in the underdoped region, below a characteristic temperature $T^*$, the symmetric solution is a free energy maximum and a pair of C4 symmetry breaking solutions, with different expectation values $\langle Q_{ij} \rangle$ and $\langle u_{ij}^2 \rangle$ in the $x$- and $y$-directed bonds (Fig. 4a), have lower free energy.

C4 symmetry breaking in the oxygen vibrator amplitudes $\left\langle u_{ij}^2\right\rangle$ is exactly the effect detected in the low-temperature STM R-plots (R is the ratio of electron to hole currents) of Ref. [6] - a key experiment in understanding cuprate physics. The STM experiment on the tunneling current into specific planar oxygens electronically detects the splitting in the van Hove singularities illustrated in Fig. 4a and in Eq. (8), and simultaneously observes the C4 splitting in the vibrational amplitude of these oxygens. Hence the experiment provides a direct critique of the interpretation of C4 symmetry-breaking in the FBM. The details of the experimental observation and its FBM interpretation are discussed in Appendix E. The FBM predicts that in the C4 symmetry-broken state the higher/lower-amplitude oxygens have filled/empty DOS peaks (Fig. 4a). Hence the higher/lower-amplitude oxygens should show as dark streaks/light spots in the R-plots, exactly as observed.

C4 symmetry breaking in the electronic structure leads to a $d$-type PG $\Delta_{ps}(\mathbf{k}) \sim \Delta_{ps}(\cos k_x - \cos k_y)/2$ (see Appendix D (D4)), where the PG, $\Delta_{ps}$, can be positive or negative in sign. The degeneracy of the saddle points at X= $(\pi, 0)$ and Y= $(0, \pi)$ is split by twice the PG, $2\Delta_{ps}$ (see Fig. 4a and Eq. (8)).

The critical condition for the existence of the C4 symmetry breaking, and hence for the existence of the pseudogap, is derived by linearizing the mean field treatment so as to obtain the conditions for instability. This analysis is detailed in Appendix F, yielding Eq. (F48). The condition can be shown to be approximately equivalent under practical conditions to

$$8K\rho(\epsilon_F) \gtrsim 1, \tag{9}$$

where $\rho(\epsilon_F)$ is the density of states at the Fermi level. The quantity $8K$ can be taken as the mean field coupling energy in the FBMII. The values of $8K$ derived from the AIMD calculations are illustrated in Table I. They are seen to be of order 1 eV. The density of states at the Fermi level is somewhat larger than 1 eV$^{-1}$ for practical doping levels, so that the pseudogap is indeed predicted to exist in the mean field theory of the FBM [8].

The FBM phase diagram for the PG is shown in Fig. 4b. The mean field result reproduces the main experimental features of the pseudogap. Daou *et al.* [4] show that the temperature boundary $T^*$ of the PG is indeed coincident with the temperature boundary of the C4 splitting, as the FBM predicts. In spatially inhomogeneous samples the spatial boundary of the PG is found to be coincident with the spatial boundary of the C4 splitting [5], again as the FBM predicts. The temperature and doping dependence of the PG seen in Fig. 4b is in reasonable agreement with experiment: at low temperature the pseudogap ranges from a maximum of about $\Delta_{ps} = 100$meV on the underdoped side, decreasing with increasing doping [5, 27–30], while $T^*$ is of about the right magnitude with the correct trend as a function of doping [1]. Note that these results are essentially the same in FBMI and FBMII, apart from the doping dependence introduced by the FBM II charge terms into Eq. (D1).

## IV. THE CDW AND FERMI SURFACE ARCS

So far we have assumed spatial uniformity of the mean field vibrational amplitude and pseudogap. What is the origin of the spatial oscillation of the C4 splitting in the oxygen vibrator amplitudes, which is observed to have a wavelength of approximately $\simeq 4$ unit cells [6]?. The observed spatial oscillation must strongly impact the C4 symmetry breaking in the electronic structure viewed in $k$-space and hence spectroscopic observations of the PG.

In a further enhancement of the model, we show in Appendix F that when the long range interaction between the oxygen charges in the $CuO_2$ plane oxygens is included - an improvement to the model we term FBMIII - then the FBM has a natural spatial charge oscillation or CDW. The new long range Coulomb piece is not important as long as only spatially uniform mean field quantitites are being considered, but it becomes significant when spatially nonuniform mean field quantities are introduced. An order of magnitude for the CDW wavevector $q_{CDW}$ derived in Appendix F is

$$q_{CDW}^2 \gtrsim \frac{4\pi e^2}{\epsilon v_c K}, \tag{10}$$

where $e$ is electronic charge, $\epsilon$ is the background dielectric constant, and $v_c$ is the unit cell volume. With $\epsilon \simeq 15$ Eq. (10) gives $\simeq 1.7$ unit cells for the CDW wavelength.

The presence of the CDW is likely to disrupt the assumed spatially uniform value for the C4 splitting, since both CDW and C4 splitting involve charge displacements on planar oxygens and are mutually coupled. A C4 splitting wave locked into the CDW is then likely to occur. The C4 splitting amplitude is expected to be optimum at some value of local charge, but the C4 splitting phase at that point can have either sign. Hence according to this argument the C4 splitting wavelength is $2\times$ the CDW wavelength. If the C4 splitting oscillation locks into the CDW at $2\times$ the CDW wavelength, then (10) is consistent with the C4 splitting wavelength being $\simeq 4$ unit cells.

The spatial oscillation of the C4 splitting implies that the sign of $\Delta_{ps}$ will vary spatially with distance $x$ parallel to the CDW wavevector $\mathbf{q}_{CDW}$, in a manner $\Delta_{ps}(\mathbf{k}) \sim \Delta_{ps}\cos(\theta)(\cos k_x - \cos k_y)/2$, where $\theta = q_{CDW}x/2$. The Fermi surface in the C4 split phase will be sensitive to the sign of $\Delta_{ps}$, and hence to the phase $\theta$. The inset in Fig. 5 shows the two FS corresponding to the limiting cases $\theta = (2n+1)\pi$ and $\theta = 2n\pi$ ($n$ = integer), and it is reasonable to assume that a nanoscopically varying order parameter will lead to FS smearing between these limits as illustrated by the shaded region in the Fig. 5 inset. A further source of spectroscopic incoherence is that the spatial oscillation of the C4 splitting is not coherent, but broken into domains [6], which has been interpreted as an effect of the nonuniform dopant distribution [31].

In $k$-space the smearing of the FS by the PG, seen in the Fig. 5 inset, is seen to be zero at the nodal line $k_x = k_y$, a manifestation of the $d$-type nature of the PG $\Delta_{ps}(\mathbf{k})$, and to increase as one goes towards the SP's at X and Y. If we make a measurement on some energy scale $E$, it is to be expected that the local FS will be well-defined at $\mathbf{k}$-points where the local PG is less than $E$, $\Delta_{ps}(\mathbf{k}) < E$, but that it will appear smeared out on energy scales where the local PG is larger than $E$, $\Delta_{ps}(\mathbf{k}) > E$. As an example of this phenomenon we could take $E$ to be temperature, $E = k_{\mathrm{B}}T$ . According to this argument, there should then be a boundary between the resolvable and unresolvable sections of the FS arc defined by $\Delta_{ps}(\mathbf{k}) = k_{\mathrm{B}}T$. In fact this FS arc effect has already been observed [32, 33]. Comparison of the data with our heuristic model in Fig. 5 is in good agreement with experiment [33], especially in the low-$T$ regime (this approach does not include the temperature-dependent quasiparticle lifetime broadening [32], which we will treat in a future paper dealing with quasiparticle dynamics).

The *ab initio* calculations we have done show that there is an underlying instability of the $CuO_2$ plane oxygens in HTS which results in the phenomenon of C4 symmetry breaking. Based on the *ab initio* calculations we derive an enhanced version of the FBM, with realistically-estimated parameters. The mean field theory of the FBM shows that C4 symmetry breaking is the underlying cause of the pseudogap, as also found by recent complementary experiments. The FBM is able to give a picture of the pseudogap phenomenology including features such as $T^*$, the doping dependence of the pseudogap, and Fermi surface arcs which are all in agreement with experiment. Including our previous success in explaining the superconductivity and doping-dependent isotope shift, we believe that the FBM approach has solid achievements in explaining the main nonmagnetic phenomena in the cuprate high temperature superconductors.

**Acknowledgments**

This work was supported by the Natural Sciences and Engineering Research Council (M.H.M.). Computational resources were provided by Sharcnet and IBM Watson's Blue Gene/L supercomputer.

## APPENDIX A: DERIVATION OF THE FBMII COUPLING

In matrix notation consider a $d$-subspace and a $p$-subspace, represented by the Hamiltonians $H^d$ and $H^p$ respectively, connected by the coupling matrix $V^{pd}$, the Hamiltonian then being,

$$H = \begin{bmatrix} H^d & V^{dp} \\ V^{pd} & H^p \end{bmatrix}. \tag{A1}$$

Projecting onto the $d$-subspace in perturbation theory

$$\widetilde{H^d} = H^d + V^{dp} \left(\epsilon_d - H^p\right)^{-1} V^{pd}. \tag{A2}$$

if $i, j$ are $d$-sites, and $l, m$ are p-orbitals

$$\widetilde{H^d}_{ij} = \epsilon_d \delta_{ij} + \sum_{l,m} V^{dp}_{il} \left(\epsilon_d - H^p\right)^{-1}_{lm} V^{pd}_{mj}. \tag{A3}$$

Now we shall neglect the $pp$ hopping matrix elements (Emery model), when $l = m$ and the $V$'s are nearest-neighbor hopping matrix elements defined as $t_{pd} > 0$ (see Fig. 3). There are 2 processes,

1. $i, j$ nearest neighbor $\langle i, j \rangle$ on the $d$-lattice, when the 2 $V$'s have opposite sign (Fig. 3)

2. $i = j$, when the 2 $V$'s have same sign

giving

$$\widetilde{H^d} = \sum_{i,\sigma} \epsilon_d n_{i\sigma} + \sum_{\langle i,j\rangle,\sigma} \frac{t^2_{p_{ij}d}}{\epsilon_{p_{ij}d}} \left(n_{i\sigma} + n_{j\sigma}\right) - \sum_{\langle i,j\rangle} \frac{t^2_{p_{ij}d}}{\epsilon_{p_{ij}d}} X_{ij}, \tag{A4}$$

where $\epsilon_{pd} = \epsilon_d - \epsilon_p > 0$ is the "oxide gap" between the oxygen $2p$ orbital energy and the higher-lying Cu $3d_{x^2-y^2}$ orbital energy, $\sigma$ is spin, $p_{ij}$ is the p-orbital between $d$-sites $i$ and $j$, and the bond order operator $X_{ij}$ is

$$X_{ij} = \sum_{\sigma} \left(c^+_{i\sigma} c_{j\sigma} + c^+_{j\sigma} c_{i\sigma}\right). \tag{A5}$$

Let us assume that the oxygen motion in some direction is $x$, and that it enters the 3-band hamiltonian via the $pd$ hopping integral

$$t_{pd} \to t_{pd} - v_{pd} x^2, \quad \text{where } v_{pd} > 0, \tag{A6}$$

then to order $v_{pd}$, and defining $t = t^2_{pd}/\epsilon_{pd}$

$$\widetilde{H^d} = (\epsilon_d + 2t) \sum_{i,\sigma} n_{i\sigma} - t \sum_{\langle i,j\rangle} X_{ij} - \frac{2t_{pd} v_{pd}}{\epsilon_{pd}} \sum_{\langle i,j\rangle,\sigma} \left(n_{i\sigma} + n_{j\sigma}\right) x^2_{ij} + \frac{2t_{pd} v_{pd}}{\epsilon_{pd}} \sum_{\langle i,j\rangle} X_{ij} x^2_{ij}. \tag{A7}$$

Restoring our original notation [8] $2t_{pd} v_{pd}/\epsilon_{pd} = v/2\sqrt{nn_s}$ ($n$ is the degeneracy of the vibrational mode, and $n_s$ is the degeneracy of the fermions, in practice $n = n_s = 2$), when the coupling $v$ is seen to be **positive**

$$\widetilde{H^d} = (\epsilon_d + 2t) \sum_{i,\sigma} n_{i\sigma} - t \sum_{\langle i,j\rangle} X_{ij} - \frac{v}{2\sqrt{nn_s}} \sum_{\langle i,j\rangle,\sigma} \left(n_{i\sigma} + n_{j\sigma}\right) x^2_{ij} + \frac{v}{2\sqrt{nn_s}} \sum_{\langle i,j\rangle} X_{ij} x^2_{ij}. \tag{A8}$$

We retrieve our previous 1-band model (next-nearest and next-next-nearest neighbor hoppings are dropped due to neglect of $t_{pp}$), but with an extra term diagonal in $d$-space. As regards the vibrator, the effect of the new term is to stiffen the vibrator with increasing hole occupation. In this respect the number operator term is dominant over the hopping term (<X> maximizes at $\simeq 0.6$).

Let us now alternatively assume that the oxygen motion enters the 3-band hamiltonian through the interaction of the electrostatic potential with the charge on the oxygen

$$\epsilon_{pd} \rightarrow \epsilon_{pd} + v_p x^2; \tag{A9}$$

where $v_p$ depends on a Madelung sum. In an ionic crystal it is arguable that the sign of $v_p$ will be positive since the environment of a negative ion typically consists of positive ions, so as the O-ion approaches them the local oxide gap $\epsilon_{pd}$ becomes larger. However in a perovskite structure the issue needs specific calculation.

Expanding to first order

$$\frac{1}{\epsilon_{pd} + v_p x^2} = \frac{1}{\epsilon_{pd}} - \frac{v_p x^2}{\epsilon_{pd}^2}. \tag{A10}$$

Returning to Eq. (A4), we insert the foregoing expansion into the 2 terms to obtain

$$\begin{aligned} \Delta \widetilde{H^d} \rightarrow & -\frac{t v_p}{\epsilon_{pd}} \sum_{\langle i,j \rangle} \left( n_{i\sigma} + n_{j\sigma} \right) x_{ij}^2 \\ & + \frac{t v_p}{\epsilon_{pd}} \sum_{\langle i,j \rangle} X_{ij} x_{ij}^2. \end{aligned} \tag{A11}$$

The effect of the oscillator correction (A11) from this mechanism can be absorbed into (A8), giving the same final result (A8) but with

$$\frac{v}{2\sqrt{n n_s}} = \left( 2 t_{pd} v_{pd} + t v_p \right) / \epsilon_{pd}. \tag{A12}$$

The sign of $v$ will be positive if the $t_{pd} v_{pd}$ term in parenthesis is dominant, or if $v_p$ is positive as argued above.

In this section we have formally derived the FBM coupling, showing the approximations involved explicitly, and demonstrated the existence of a new term in the coupling, extending the initial FBM [8], termed FBMI, to the model including charge coupling, the FBMII.

## APPENDIX B: THE COMPLETE FBMII HAMILTONIAN

The FBMII Hamiltonian involves three pieces

$$H = H^v + H^e + H^{ev}. \tag{B1}$$

In $H$ the Cu sites, which define the unit cell, are defined as 2D integral-component vectors $\mathbf{i} = (i_x, i_y)$ (lattice constant is taken as unity). The two oxygens in each unit cell $\mathbf{i}$ are located at the sites $\mathbf{i}+\widehat{\alpha}/2$, where $\widehat{\alpha}$ is a unit vector along the $x$- or $y$- axes, hence $\widehat{\alpha}$ defines whether the oxygen is in a Cu-O-Cu bond oriented along the $x$- or $y$- direction.

In the vibrator piece $H^v$ the oxygen degree of freedom is an $n$-component vector $\mathbf{x}_{\mathbf{i}+\widehat{\alpha}/2}$, where $n = 1$ if a single mode is dominant (as assumed in the manuscript), $n = 2$ if the two modes transverse to the Cu-O-Cu bond are roughly equivalent, or in a case now considered unlikely (as the along-bond mode is found to be weakly coupled) $n = 3$ if the two transverse modes and the along-bond mode can all be considered equivalent. $H^v$ is given by

$$H^v = \sum_{\mathbf{i},\alpha=\mathbf{x}}^{\mathbf{y}} \left[ \frac{1}{2m} p^2_{\mathbf{i}+\widehat{\alpha}/2} + \frac{\chi_0}{2} x^2_{\mathbf{i}+\widehat{\alpha}/2} + \frac{w}{8n} \left( x^2_{\mathbf{i}+\widehat{\alpha}/2} \right)^2 \right]. \tag{B2}$$

In $H^v$ the scalar products $\mathbf{x}_{\mathbf{i}+\widehat{\alpha}/2} \cdot \mathbf{x}_{\mathbf{i}+\widehat{\alpha}/2}$ are abbreviated to $x^2_{\mathbf{i}+\widehat{\alpha}/2}$, and a momentum $\mathbf{p}_{\mathbf{i}+\widehat{\alpha}/2}$ conjugate to coordinate $\mathbf{x}_{\mathbf{i}+\widehat{\alpha}/2}$ is introduced, to define the vibrator kinetic energy, with $m$ the oxygen mass ($M$ in the Ms.). The "bare" bond force constant is $\chi_0$. The quartic term, with coefficient $w$, is assumed in the degenerate case to be radially ($n = 2$) or spherically ($n = 3$) symmetric.

The electronic piece $H^e$ is

$$H^e = -\frac{1}{2} \sum_{\mathbf{i},\mathbf{j},\sigma} t\left(\mathbf{i} - \mathbf{j}\right) c^+_{\mathbf{i},\sigma} c_{\mathbf{j},\sigma}, \tag{B3}$$

where $c^+_{\mathbf{i},\sigma}$ ($c_{\mathbf{i},\sigma}$) denote respectively the creation (destruction) operators for the $3d_{x^2-y^2}$ orbital (or, more rigorously, the $d_{x^2-y^2}$-type Cu$3d$-O$2p$ antibonding Wannier function) on lattice site $\mathbf{i}$ of spin $\sigma$. The strongest interaction is the nearest neighbor hopping integral $t(\pm1, 0) = t(0, \pm1) = t$, ($t$ is positive), followed by the next-nearest neighbor interaction $t(\pm1, \pm1) = t'$, ($t'$ is negative) and then the 3rd-nearest neighbor interaction $t(\pm2, 0) = t(0, \pm2) = t''$ ($t''$ is positive). The band eigenvalues $\epsilon_{\mathbf{k}}$ of (B3) are

$$\begin{aligned} \epsilon_{\mathbf{k}} = &-2t(\cos k_x + \cos k_y) - 4t' \cos k_x \cos k_y \\ &- 2t''(\cos 2k_x + \cos 2k_y). \end{aligned} \tag{B4}$$

The model band structure has a minimum at $\Gamma$ ($\mathbf{k} = (0, 0)$), a maximum at $Z$ ($\mathbf{k} = (\pi, \pi)$), and saddle points (SP) at X ($\mathbf{k} = (0, \pi)$), and Y ($\mathbf{k} = (\pi, 0)$). As a result of the saddle points, located at $\epsilon_{\mathbf{SP}} = 4t' - 4t''$, the density of states (DOS) has a logarithmic peak (van

Hove singularity or vHs) at $\epsilon_{\mathbf{SP}}$ which is found from ARPES and band structure calculations for near-optimally doped systems to lie close to the Fermi level [24, 34] - the resulting high DOS at the Fermi level strongly enhances the FBM coupling. The total band width is $8t$.

The electron-vibrator coupling piece is

$$H^{ev} = \frac{-v}{2\sqrt{nn_s}} \sum_{\mathbf{i},\alpha=\mathbf{x}}^{\mathbf{y}} x^2_{\mathbf{i}+\hat{\alpha}/2} \times \left[ \sum_\sigma \left( n_{\mathbf{i},\sigma} + n_{\mathbf{i}+\hat{\alpha},\sigma} \right) - X_{\mathbf{i}+\hat{\alpha}/2} \right] ; \tag{B5}$$

$$X_{\mathbf{i}+\hat{\alpha}/2} = \sum_\sigma \left( c^+_{\mathbf{i},\sigma} c_{\mathbf{i}+\hat{\alpha},\sigma} + c^+_{\mathbf{i}+\hat{\alpha},\sigma} c_{\mathbf{i},\sigma} \right) , \tag{B6}$$

where the bond order operator $X$ is associated with the oxygen site at the bond center, and we have defined in the mixed degeneracy factor $(nn_s)^{-1/2}$, where $n_s = 2$ is the spin degeneracy, to make the term of order $\sqrt{nn_s}$, motivated by a version of large-$N$ theory jointly expanding in $1/n$ and $1/n_s$ [8]. In Ref. [8] only the $X$-piece of (B5) was included, a level termed FBMI.

The combination $\sum_\sigma \left( n_{\mathbf{i},\sigma} + n_{\mathbf{i}+\hat{\alpha},\sigma} \right) - X_{\mathbf{i}+\hat{\alpha}/2}$ can also be written in more compact form, defining the antibonding orbital $|a, \mathbf{i}+\hat{\alpha}/2\rangle = (|\mathbf{i}\rangle - |\mathbf{i}+\hat{\alpha}\rangle)/\sqrt{2}$, with number operator (summing over spin) denoted $Q_{\mathbf{i}+\hat{\alpha}/2} = \sum_\sigma c^+_{a,\mathbf{i}+\hat{\alpha}/2,\sigma} c_{a,\mathbf{i}+\hat{\alpha}/2,\sigma}$.

$$\frac{1}{2} \left( \sum_\sigma (n_{\mathbf{i},\sigma} + n_{\mathbf{i}+\hat{\alpha},\sigma}) - X_{\mathbf{i}+\hat{\alpha}/2} \right) = Q_{\mathbf{i}+\hat{\alpha}/2}. \tag{B7}$$

The complete Hamiltonian $H = H^v + H^e + H^{ev}$ is then

$$H = \sum_{\mathbf{i},\alpha=\mathbf{x}}^{\mathbf{y}} \left[ \frac{1}{2m} p^2_{\mathbf{i}+\hat{\alpha}/2} + \frac{\chi_0}{2} x^2_{\mathbf{i}+\hat{\alpha}/2} + \frac{w}{8n} \left( x^2_{\mathbf{i}+\hat{\alpha}/2} \right)^2 \right] - \frac{1}{2} \sum_{\mathbf{i},\mathbf{j},\sigma} t(\mathbf{i}-\mathbf{j})\, c^+_{\mathbf{i},\sigma} c_{\mathbf{j},\sigma} - \frac{v}{\sqrt{nn_s}} \sum_{\mathbf{i},\alpha=\mathbf{x}}^{\mathbf{y}} x^2_{\mathbf{i}+\hat{\alpha}/2} Q_{\mathbf{i}+\hat{\alpha}/2}. \tag{B8}$$

Note that in Eq.(B8) $K = v^2/w$ defines a coupling energy.

## APPENDIX C: DETERMINATION OF COUPLING $v$

Calculation of the oxygen PE surface as a function of doping is not an ideal approach to calculationg the FBM coupling constant. The coupling in the FBM Hamiltonian is to the number of electrons $Q$ in the antibonding orbital, which mainly involves states at the top of the $d$-band and will be filled mainly by adding electrons rather than, on the contrary, holes as was done (for reasons of computational stability) in Fig. 2 of the Ms..

The method adopted to calculate the coupling strength $v$ is based on comparing the shift in band structure energies when the oxygen location is perturbed with the same shift deduced from the FBM Hamiltonian. The FBM coupling (third term in Eq.(B8)) leads to splittings in the tight-binding band structure. If all oxygens in the $x$-oriented bonds are globally shifted by $u_x$, and all oxygens in the $y$-oriented bonds by $u_y$, there is a splitting of the band energy between the band energy $\epsilon_X$ at the saddle point (SP) X= $(\pi, 0)$, and $\epsilon_Y$ at Y= $(0, \pi)$, given by $\epsilon_X - \epsilon_Y = \sqrt{2/n}v\left(u_x^2 - u_y^2\right)$. By numerically calculating the band structure with first the $x$-oxygens displaced, and then the $y$-oxygens, and subtracting the corresponding band structure energies energies at, say, the SP X, any isotropic shift resulting from displacing a single oxygen can be cancelled out and the coupling $v$ determined. The results are shown in Table I.

## APPENDIX D: MEAN FIELD APPROXIMATION

Mean field theory is a useful step in investigating the properties of many models. In the FBM, the mean field approximation decouples the electronic and vibrational parts of the Hamiltonian. In the vibrational part, an expectation value of the electronic terms shifts the oscillator harmonic frequency, the expectation value being assumed spatially uniform, but it can be different in the $x$- and $y$- bonds (in this section we return to the notation in the Ms.):

$$H^{vib} = \sum_{\langle i,j \rangle} \frac{p_{ij}^2}{2M} + \frac{1}{2}\sum_{\langle i,j \rangle} \chi_0 u_{ij}^2 + \frac{w}{8}\sum_{\langle i,j \rangle} u_{ij}^4 + \frac{v}{2\sqrt{2}} \sum_{\langle i,j \rangle,\sigma} \left(2 - 2p + \left\langle c_{i,\sigma}^+ c_{j,\sigma} + c_{j,\sigma}^+ c_{i,\sigma} \right\rangle\right) u_{ij}^2. \tag{D1}$$

$H^{vib}$ can easily be diagonalized in a harmonic oscillator basis. In the electronic part, the expectation value of the square of the oscillator amplitude has been taken,

$$H^{el} = \sum_{\mathbf{k},\sigma} \epsilon_{\mathbf{k}} n_{\mathbf{k},\sigma} + \frac{v}{2\sqrt{2}} \sum_{\langle i,j \rangle,\sigma} \left[c_{i,\sigma}^+ c_{j,\sigma} + c_{j,\sigma}^+ c_{i,\sigma}\right] \left\langle u_{ij}^2 \right\rangle, \tag{D2}$$

giving a band structure problem in which there are new nearest-neighbor hopping terms $\left(v/2\sqrt{2}\right) \left[c_{i,\sigma}^+ c_{j,\sigma} + c_{j,\sigma}^+ c_{i,\sigma}\right] \left\langle u_{ij}^2 \right\rangle$ (the uniform shift represented by the number operator terms does not change the band structure and is omitted) with the effect of reducing the nearest-neighbor hopping integral. Allowing the oscillator amplitude squared for the $x$-directed $\left\langle u_{ij}^2 \right\rangle_x$ and $y$-directed $\left\langle u_{ij}^2 \right\rangle_y$ bonds to be unequal (the C4 symmetry-split case), the band structure is changed to

$$\widetilde{\epsilon}_{\mathbf{k}} = \epsilon_{\mathbf{k}} + \frac{v}{\sqrt{2}} \left\langle u_{ij}^2 \right\rangle_x \cos k_x + \frac{v}{\sqrt{2}} \left\langle u_{ij}^2 \right\rangle_y \cos k_y. \tag{D3}$$

Using the band structure $\widetilde{\epsilon}_{\mathbf{k}}$ (D3) the expectation values $\left\langle c_{i,\sigma}^{+} c_{j,\sigma} + c_{j,\sigma}^{+} c_{i,\sigma} \right\rangle$ for $x$-oriented and $y$-oriented bonds are calculated, hence defining two quartic Hamiltonians (D1) whose exact solution yields the squared vibrator amplitudes $\left\langle u_{ij}^2 \right\rangle_x$ and $\left\langle u_{ij}^2 \right\rangle_y$. These interconnected electronic and quartic problems are then solved self-consistently as regards the expectation values. The parameters used were similar to Table I, $v = 0.0198$ au, $w = 0.085$ au, the oscillator bare force constant was $\chi_0 = -0.0225$ au. The band structure is parametrized by the (negative of the) hopping matrix elements, the nearest-neighbor hopping matrix element $t = 0.25$ eV, next-nearest-neighbor hopping m.e. $t' = -0.05$ eV, and third next-nearest-neighbor hopping m.e. $t'' = 27.2$ meV.

We can rewrite the effective band structure as

$$\widetilde{\epsilon}_{\mathbf{k}} = \overline{\epsilon_{\mathbf{k}}} + \frac{1}{2} \Delta_{ps} \left( \cos k_x - \cos k_y \right), \tag{D4}$$

where $\Delta_{ps} = \left( v/\sqrt{2} \right) \left( \left\langle u_{ij}^2 \right\rangle_x - \left\langle u_{ij}^2 \right\rangle_y \right)$ is the pseudogap, and the renormalized nearest-neighbor hopping $\left( v/2\sqrt{2} \right) \left( \left\langle u_{ij}^2 \right\rangle_x + \left\langle u_{ij}^2 \right\rangle_y \right)$ is absorbed into $\overline{\epsilon_{\mathbf{k}}}$. The experimental data [6] show that the pseudogap is not uniform over the sample as we have, for simplicity, assumed, but the coherence length over which the sign of $\Delta_{ps}^0$ varies is quite short, only a few lattice spacings. Probably as a result of this nanoscopic domain structure, the phase boundary of the pseudogap region is not typically found experimentally to constitute a true, sharp, phase boundary [1].

The variation of pseudogap with doping at low temperature seen in the contour plot (Fig. 4b) is similar to that seen in experimental data [30].

## APPENDIX E: INTENSITY VARIATION IN EXPERIMENTAL $R$-PLOTS

In order to model the experimental behavior in the STM experiments [6] on C4 symmetry-split systems, we calculated the projected DOS for a 3-band model with the basis of oxygen $2p_x$ and $2p_y$ orbitals and Cu $3d_{x^2-y^2}$ orbitals shown in Fig. 3. The $pd\sigma$ hopping matrix element is $t_{pd} = 1.12$ eV. There are $pp$ hopping matrix elements between nearest-neighbor $2p_x$ and $2p_y$ orbitals given by $t_{pp} = -0.528$ eV, and an oxide gap $\epsilon_d - \epsilon_p = 6$ eV. A spatially-uniform pseudogap is introduced by modifying the $t_{pd}$ matrix elements to $t_{p_xd} = t_{pd} + \Delta t$ (i.e. for the lower vibrational amplitude oxygen) and $t_{p_yd} = t_{pd} - \Delta t$ (i.e. for the higher vibrational amplitude oxygen), where $\Delta t = 0.0375$ eV (the argument below only depends on these being semiquantitatively correct).

The results for the DOS projected into the oxygen $2p_x$ orbitals (lying in $x$-oriented Cu-O-Cu bonds - see Fig. 3) and oxygen $2p_y$ orbitals are different, as seen in Fig. 4a. The DOS peak associated with the van Hove singularity is seen in Fig. 4a to be split, the peak above the Fermi level being localized only on the lower vibrational amplitude oxygen, and the peak below the Fermi level being localized only on the higher vibrational amplitude oxygen. The STM $R$-map technique [6] for detecting the C4 splitting experimentally involves the ratio $R$ of the tunneling current into the empty DOS to the hole current into the filled DOS. Evidently from Fig. 4, $R$ is predicted to be large on the low amplitude oxygens and small on the high amplitude oxygens, in agreement with the observation [6], in which the high amplitude oxygens are associated with dark streaks in the $R$-map, while the low amplitude oxygens are associated with bright spots. Note that the C4 splitting is characterized by nanoscale domains [6], rather than being spatially uniform as assumed in the Fig. 4a calculations.

## APPENDIX F: THE FBM HAMILTONIANAN WITH LONG-RANGE INTERACTION IN MEAN FIELD

### 1. The FBM Hamiltonianan with Long-Range Coulomb Interaction

Including the Long-Range Coulomb Interaction (LRCI) between the charges $Q$ in the antibonding orbital in (B8) gives the Hamiltonian

$$H=\sum_{\mathbf{i},\alpha=\mathbf{x}}^{\mathbf{y}}\left[\frac{1}{2m}p^2_{\mathbf{i}+\widehat{\alpha}/2}+\frac{\chi_0}{2}x^2_{\mathbf{i}+\widehat{\alpha}/2}+\frac{w}{8n}\left(x^2_{\mathbf{i}+\widehat{\alpha}/2}\right)^2\right]+\sum_{\mathbf{k},\sigma}\epsilon^0_{\mathbf{k}}n_{\mathbf{k},\sigma} \tag{F1}$$
$$-\frac{v}{\sqrt{nn_s}}\sum_{\mathbf{i},\alpha=\mathbf{x}}^{\mathbf{y}}x^2_{\mathbf{i}+\widehat{\alpha}/2}Q_{\mathbf{i}+\widehat{\alpha}/2}$$
$$+\frac{e^2}{\epsilon n_s}\sum_{\mathbf{i},\mathbf{j},\alpha,\beta}Q_{\mathbf{i}+\widehat{\alpha}/2}\frac{\int d^3r\int d^3r' f_\alpha\left(\mathbf{r}-\mathbf{r}_{\mathbf{i}+\widehat{\alpha}/2}\right)f_\beta\left(\mathbf{r}'-\mathbf{r}_{\mathbf{j}+\widehat{\beta}/2}\right)}{|\mathbf{r}-\mathbf{r}'|}Q_{\mathbf{j}+\widehat{\beta}/2}.$$

Here $f_x(\mathbf{r}-\mathbf{r}_{\widehat{\mathbf{x}}/\mathbf{2}})$, $f_y(\mathbf{r}-\mathbf{r}_{\widehat{\mathbf{y}}/\mathbf{2}})$ are the form factors (charge probability distribution) of the charges in the $x$- and $y$- antibonding orbitals on a bond from the origin to $\widehat{\mathbf{x}}$, $\widehat{\mathbf{y}}$. $f_\alpha\left(\mathbf{r}-\mathbf{r}_{\widehat{\alpha}/2}\right)$ can be written in the single band basis as

$$f_\alpha\left(\mathbf{r}-\mathbf{r}_{\widehat{\alpha}/2}\right)=\frac{1}{2}\left(\psi_{3d}(\mathbf{r})-\psi_{3d}(\mathbf{r}-\mathbf{r}_{\widehat{\alpha}})\right)^2. \tag{F2}$$

$f_\alpha(\mathbf{r})$ is assumed normalized to unity $\int d^3r f_\alpha(\mathbf{r})=1$, as will be the case if the $3d$-orbitals on adjacent sites are orthonormal. $\epsilon$ is a background dielectric constant of order several. A factor of 2 has been incorporated into the LRCI in order that it be correct for $n_s=2$.

This can be written more compactly by defining the LRCI $2\times 2$ tensor

$$V_{\alpha\beta}(\mathbf{r}_{\mathbf{i}+\widehat{\alpha}/2}-\mathbf{r}_{\mathbf{j}+\widehat{\beta}/2})=\frac{2e^2}{\epsilon}\int d^3r\int d^3r'\frac{f_\beta\left(\mathbf{r}-\mathbf{r}_{\mathbf{i}+\widehat{\alpha}/2}\right)f_\gamma\left(\mathbf{r}'-\mathbf{r}_{\mathbf{j}+\widehat{\beta}/2}\right)}{|\mathbf{r}-\mathbf{r}'|}, \tag{F3}$$

$$H=\sum_{\mathbf{i},\alpha=\mathbf{x}}^{\mathbf{y}}\left[\frac{1}{2m}p^2_{\mathbf{i}+\widehat{\alpha}/2}+\frac{\chi_0}{2}x^2_{\mathbf{i}+\widehat{\alpha}/2}+\frac{w}{8n}\left(x^2_{\mathbf{i}+\widehat{\alpha}/2}\right)^2\right]+\sum_{\mathbf{k},\sigma}\epsilon^0_{\mathbf{k}}n_{\mathbf{k},\sigma} \tag{F4}$$
$$-\frac{v}{\sqrt{nn_s}}\sum_{\mathbf{i},\alpha=\mathbf{x}}^{\mathbf{y}}x^2_{\mathbf{i}+\widehat{\alpha}/2}Q_{\mathbf{i}+\widehat{\alpha}/2}+\frac{1}{2n_s}\sum_{\mathbf{i},\mathbf{j},\alpha,\beta}Q_{\mathbf{i}+\widehat{\alpha}/2}V_{\alpha\beta}(\mathbf{r}_{\mathbf{i}+\widehat{\alpha}/2}-\mathbf{r}_{\mathbf{j}+\widehat{\beta}/2})Q_{\mathbf{j}+\widehat{\beta}/2}.$$

### 2. FBM with LRCI Hamiltonian in Mean Field

In this section we employ a mean field formulation which allows mean field quantities to vary in space, but does so in a linearized regime where the spatially-varying quantities are small. Hence the treatment is valid near the phase boundary where the spatially-varying quantities become nonzero.

In mean field approximation, the mean field Hamiltonian is (to within a constant)

$$H_{mf} = \sum_{\mathbf{i},\alpha=\mathbf{x}}^{\mathbf{y}} \left[ \frac{1}{2m} p^2_{\mathbf{i}+\hat{\alpha}/2} + \frac{\chi_{\mathbf{i}+\hat{\alpha}/2}}{2} x^2_{\mathbf{i}+\hat{\alpha}/2} \right] \tag{F5}$$

$$+ \sum_{\mathbf{k},\sigma} \epsilon^0_{\mathbf{k}} n_{\mathbf{k},\sigma} + \sum_{\mathbf{i},\alpha=\mathbf{x}}^{\mathbf{y}} \eta_{\mathbf{i}+\hat{\alpha}/2} Q_{\mathbf{i}+\hat{\alpha}/2},$$

where we defined complete oscillator stiffness $\chi_{\mathbf{i}+\hat{\alpha}/2}$ and complete bond "potential" $\eta_{\mathbf{i}+\hat{\alpha}/2}$ as

$$\chi_{\mathbf{i}+\hat{\alpha}/2} = \chi_0 + \frac{w}{2n} \left\langle x^2_{\mathbf{i}+\hat{\alpha}/2} \right\rangle - \frac{2v}{\sqrt{nn_s}} \left\langle Q_{\mathbf{i}+\hat{\alpha}/2} \right\rangle, \tag{F6}$$

$$\eta_{\mathbf{i}+\hat{\alpha}/2} = \frac{-v}{\sqrt{nn_s}} \left\langle x^2_{\mathbf{i}+\hat{\alpha}/2} \right\rangle \tag{F7}$$

$$+ \frac{1}{n_s} \sum_{\mathbf{j},\beta} V_{\alpha\beta}(\mathbf{r}_{\mathbf{i}+\hat{\alpha}/2} - \mathbf{r}_{\mathbf{j}+\hat{\beta}/2}) \left\langle Q_{\mathbf{j}+\hat{\beta}/2} \right\rangle .$$

Here we are exploiting the fact that the Coulomb potential and the (oscillator amplitude)$^2$ interact with the bond charge $Q$ in the same way.

It is useful to distinguish quantities nonuniform in space, which will be prefixed with $\Delta$, and spatial averages denoted with a bar

$$H_{mf} = \sum_{\mathbf{i},\alpha=\mathbf{x}}^{\mathbf{y}} \left[ \frac{1}{2m} p^2_{\mathbf{i}+\hat{\alpha}/2} + \frac{1}{2} \left( \chi_0 + \Delta\chi_{\mathbf{i}+\hat{\alpha}/2} \right) x^2_{\mathbf{i}+\hat{\alpha}/2} \right]$$

$$+ \sum_{\mathbf{k},\sigma} \epsilon_{\mathbf{k}} n_{\mathbf{k},\sigma} + \sum_{\mathbf{i},\alpha=\mathbf{x}}^{\mathbf{y}} \Delta\eta_{\mathbf{i}+\hat{\alpha}/2} Q_{\mathbf{i}+\hat{\alpha}/2}, \overline{\chi} \tag{F8}$$

$$= \chi_0 + \frac{w}{2n} \overline{\langle x^2 \rangle} - \frac{2v}{\sqrt{nn_s}} \overline{\langle Q \rangle}; \tag{F9}$$

$$\Delta\chi_{\mathbf{i}+\hat{\alpha}/2} = \frac{w}{2} \left\langle \Delta x^2_{\mathbf{i}+\hat{\alpha}/2} \right\rangle - \frac{2v}{\sqrt{nn_s}} \left\langle \Delta Q_{\mathbf{i}+\hat{\alpha}/2} \right\rangle, \tag{F10}$$

$$\Delta\eta_{\mathbf{i}+\hat{\alpha}/2} = \frac{-v}{\sqrt{nn_s}} \left\langle \Delta x^2_{\mathbf{i}+\hat{\alpha}/2} \right\rangle \tag{F11}$$

$$+ \frac{1}{n_s} \sum_{\mathbf{j},\beta} V_{\alpha\beta}(\mathbf{r}_{\mathbf{i}+\hat{\alpha}/2} - \mathbf{r}_{\mathbf{j}+\hat{\beta}/2}) \left\langle \Delta Q_{\mathbf{j}+\hat{\beta}/2} \right\rangle .$$

Here $\epsilon_{\mathbf{k}}$ is understood to include the $\overline{\eta}$ effects, and electrostatic effects are assumed zero in the uniform system which is site-neutral.

### 3. Linearize Vibrator Response

We shall linearize the response $\left\langle x^2_{\mathbf{i}+\widehat{\alpha}/2}\right\rangle$ of the Einstein vibrator on site $\mathbf{i}+\widehat{\alpha}/2$ to changes in the vibrator stiffness $\chi_{\mathbf{i}+\widehat{\alpha}/2}$,

$$\left\langle \Delta x^2_{\mathbf{i}+\widehat{\alpha}/2}\right\rangle = \left\langle x^2_{\mathbf{i}+\widehat{\alpha}/2}\right\rangle - \overline{\left\langle x^2\right\rangle} \tag{F12}$$

$$= -An\Delta\chi_{\mathbf{i}+\widehat{\alpha}/2} \tag{F13}$$

$$= -An\left(\chi_{\mathbf{i}+\widehat{\alpha}/2} - \overline{\chi}\right), \tag{F14}$$

where

$$A = \frac{\hbar}{4m^2\overline{\omega}^3} g(\frac{\hbar\overline{\omega}}{2kT}); \tag{F15}$$

$$g(x) = \coth(x) + \frac{x}{\sinh^2(x)}; \quad m\overline{\omega}^2 = \overline{\chi}. \tag{F16}$$

So

$$\left\langle \Delta x^2_{\mathbf{i}+\widehat{\alpha}/2}\right\rangle = -An\left(\frac{w}{2n}\left\langle \Delta x^2_{\mathbf{i}+\widehat{\alpha}/2}\right\rangle - \frac{2v}{\sqrt{nn_s}}\left\langle \Delta Q_{\mathbf{i}+\widehat{\alpha}/2}\right\rangle\right), \tag{F17}$$

or

$$\left\langle \Delta x^2_{\mathbf{i}+\widehat{\alpha}/2}\right\rangle = \frac{2v\widetilde{A}\sqrt{n}}{\sqrt{n_s}}\left\langle \Delta Q_{\mathbf{i}+\widehat{\alpha}/2}\right\rangle, \text{ where} \tag{F18}$$

$$\widetilde{A} = \frac{A}{\left(1 + A\frac{w}{2}\right)}. \tag{F19}$$

### 4. Electronic linear response

The assumption here is that we are near $T^*$, hence in all channels the electronic system can be assumed to have a linear response

$$\left\langle \Delta Q_{\mathbf{i}+\widehat{\alpha}/2}\right\rangle = -n_s \sum_{\mathbf{j},\beta} R^{\alpha\beta}_{\mathbf{i}+\widehat{\alpha}/2-,\mathbf{j}-\widehat{\beta}/2}\Delta\eta_{\mathbf{j}+\widehat{\beta}/2} \tag{F20}$$

$$= -n_s \sum_{\mathbf{j},\beta} R^{\alpha\beta}_{\mathbf{i}+\widehat{\alpha}/2-,\mathbf{j}-\widehat{\beta}/2}\left(\frac{-v}{\sqrt{nn_s}}\left\langle \Delta x^2_{\mathbf{j}+\widehat{\beta}/2}\right\rangle + \frac{1}{n_s}\sum_{\mathbf{k},\gamma} V_{\beta\gamma}(\mathbf{r}_{\mathbf{j}+\widehat{\beta}/2} - \mathbf{r}_{\mathbf{k}+\widehat{\gamma}/2})\left\langle \Delta Q_{\mathbf{k}+\widehat{\gamma}/2}\right\rangle\right), \tag{F21}$$

where $R_{\mathbf{i}+\widehat{\alpha}/2,\mathbf{j}+\widehat{\beta}/2}$ is a $QQ$ response function. Using (F18) this can be written in the electronic space

$$\begin{aligned}
&\left\langle \Delta Q_{\mathbf{i}+\widehat{\alpha}/2}\right\rangle - 2\widetilde{K}\sum_{\mathbf{j},\beta} R^{\alpha\beta}_{\mathbf{i}+\widehat{\alpha}/2-,\mathbf{j}-\widehat{\beta}/2}\left\langle \Delta Q_{\mathbf{j}+\widehat{\beta}/2}\right\rangle \\
&= -\sum_{\mathbf{j},\beta} R^{\alpha\beta}_{\mathbf{i}+\widehat{\alpha}/2-,\mathbf{j}-\widehat{\beta}/2} \\
&\times \sum_{\mathbf{k},\gamma} V_{\beta\gamma}(\mathbf{r}_{\mathbf{j}+\widehat{\beta}/2} - \mathbf{r}_{\mathbf{k}+\widehat{\gamma}/2})\left\langle \Delta Q_{\mathbf{k}+\widehat{\gamma}/2}\right\rangle,
\end{aligned} \tag{F22}$$

where we have introduced the effective interaction

$$\widetilde{K} = v^2\widetilde{A} = K\frac{Aw}{1+\frac{1}{2}Aw}; \quad K = \frac{v^2}{w}. \tag{F23}$$

The foregoing equation is now a homogeneous linear equation in the discrete variables $\left\langle \Delta Q^{\alpha}_{\mathbf{i}+\widehat{\alpha}/2}\right\rangle$.

Writing the linear equation (F22) as

$$\left\langle \Delta Q_{\mathbf{i}+\widehat{\alpha}/2}\right\rangle = 2\widetilde{K}\sum_{\mathbf{j},\beta} R^{\alpha\beta}_{\mathbf{i}+\widehat{\alpha}/2-,\mathbf{j}-\widehat{\beta}/2}\left\langle \Delta Q_{\mathbf{j}+\widehat{\beta}/2}\right\rangle - \sum_{\mathbf{j},\beta} R^{\alpha\beta}_{\mathbf{i}+\widehat{\alpha}/2-,\mathbf{j}-\widehat{\beta}/2}\sum_{\mathbf{k},\gamma} V_{\beta\gamma}(\mathbf{r}_{\mathbf{j}+\widehat{\beta}/2} - \mathbf{r}_{\mathbf{k}+\widehat{\gamma}/2})\left\langle \Delta Q_{\mathbf{k}+\widehat{\gamma}/2}\right\rangle, \tag{F24}$$

the LHS is the response of the bond charge $Q$ to the 2 terms on the RHS. The first term on the RHS is the bond-local response of the vibrator to the local bond charge, which then produces a contribution to the bond charge elsewhere via the nonlocal electronic response. The second term on the RHS is the nonlocal effect of the Coulomb potential produced by remote bond charges, on the potential in a given bond, which then produces a contribution to the bond charge elsewhere via the nonlocal electronic response. The response produced by coupling through the vibrator is attractive (a pairing interaction) and that via the Coulomb interaction is of course repulsive.

### 5. Response Functions for $q = 0$

The essence of the long-wavelength behavior of the RF's can be obtained by looking at the uniform limit. Define the sum over space of the $\alpha$-bond charge

$$Q_\alpha = \frac{1}{2}\sum_{\mathbf{i},\sigma}\left(n_{\mathbf{i}-\widehat{\alpha}/2,\sigma} + n_{\mathbf{i}+\widehat{\alpha}/2,\sigma}\right)$$
$$-\frac{1}{2}\sum_{\mathbf{i},\sigma}\left(c^+_{\mathbf{i}-\widehat{\alpha}/2,\sigma}c_{\mathbf{i}+\widehat{\alpha}/2,\sigma} + c^+_{\mathbf{i}+\widehat{\alpha}/2,\sigma}c_{\mathbf{i}-\widehat{\alpha}/2,\sigma}\right) \tag{F25}$$

where we refer to bond center as origin of bond. Rewriting in $k$-space

$$Q_\alpha = \sum_{\mathbf{k},\sigma} n_{\mathbf{k},\sigma} - \sum_{\mathbf{k},\sigma}\cos\left(k_\alpha\right) n_{\mathbf{k},\sigma} = \sum_{\mathbf{k},\sigma}\left(1-\cos\left(k_\alpha\right)\right) n_{\mathbf{k},\sigma}. \tag{F26}$$

The expectation value of $Q_\alpha$ is

$$\langle Q_\alpha\rangle = \sum_{\mathbf{k},\sigma}\left(1-\cos\left(k_\alpha\right)\right)\langle n_{\mathbf{k},\sigma}\rangle \tag{F27}$$
$$= n_s\sum_{\mathbf{k}}\left(1-\cos\left(k_\alpha\right)\right) f\left(\epsilon_{\mathbf{k}}-\mu\right), \tag{F28}$$

where $f$ is the Fermi function. $\langle Q_\alpha\rangle$ is positive, as is correct for the occupation number of the $\alpha$-oriented bond antibonding orbital.

To get the RF (F20) we need to differentiate with respect to changing the quantities in $H - \mu N$ by changing the coefficients of the two parts of $Q_\beta$. The coefficient of the number operator is $-\mu$. The coefficient of the second term in (F25) is $(-)$ the hopping integral $t$, though only $1/2$ is changed by $Q_\beta$, so

$$R^{\alpha\beta} = -\sum_{\mathbf{k}}\left(1-\cos\left(k_\alpha\right)\right) f'\left(\epsilon_{\mathbf{k}}\right)$$
$$\times\left[\frac{\partial\left(\epsilon_{\mathbf{k}}-\mu\right)}{\partial-\mu} - \frac{1}{2}\frac{\partial\left(\epsilon_{\mathbf{k}}-\mu\right)}{\partial t_\beta}\right] \tag{F29}$$
$$R^{\alpha\beta} = -\sum_{\mathbf{k}}\left(1-\cos\left(k_\alpha\right)\right)\left(1-\cos\left(k_\beta\right)\right) f'\left(\epsilon_{\mathbf{k}}\right). \tag{F30}$$

Note that at low temperatures $f'\left(\epsilon_{\mathbf{k}}\right) = -\delta\left(\epsilon_{\mathbf{k}}-\mu\right)$, so that the RF's are weighted DOS's at the Fermi level. The weighting will be dominated by the saddle points at X$=(\pi,0)$ and Y$=(0,\pi)$. These points contribute, X to $R^{xx}$ and Y to $R^{yy}$, but neither X or Y contributes to $R^{xy}$ or $R^{yx}$. The RF is always **positive**, but because the off-diagonal terms miss out on the SP contribution, they are expected to be smaller, hence we write

$$R^{xx} = R^{yy} = R^{>}, \qquad \text{(F31)}$$
$$R^{xy} = R^{yx} = R^{<}.$$

Numerical work suggests that the off-diagonal elements of $R$ are as much as an order of magnitude lower than the diagonal terms.

### 6. Fourier Transform

Spatial FT's are defined by

$$\overline{f}\,(\mathbf{q}) = \sum_{\mathbf{i}} e^{i\mathbf{q}.\mathbf{r_i}} f\,(\mathbf{r_i})\,; \quad q_\alpha = \frac{2\pi n_\alpha}{N_\alpha} \qquad \text{(F32)}$$

$$\sum_{\mathbf{i}} e^{i\mathbf{q}.\mathbf{r_i}} = N\delta_{\mathbf{q},\mathbf{0}}; \quad N = \Pi_\alpha N_\alpha; \qquad \text{(F33)}$$

$$f\,(\mathbf{r_i}) = \frac{1}{N}\sum_{\mathbf{q}} e^{-i\mathbf{q}.\mathbf{r_i}} \overline{f}\,(\mathbf{q})\,; \qquad \text{(F34)}$$

Applying the FT's we get a $2 \times 2$ equation for the FT of $\Delta Q$:

$$Q_\alpha(\mathbf{q}) - 2\widetilde{K}\sum_{\beta} R^{\alpha\beta}(\mathbf{q})Q_\beta(\mathbf{q})$$
$$= -\sum_{\beta,\gamma} R^{\alpha\gamma}(\mathbf{q})V_{\gamma\beta}(\mathbf{q})Q_\beta(\mathbf{q}), \qquad \text{(F35)}$$

where

$$Q_\alpha(\mathbf{q}) = \sum_{\mathbf{i}} \left\langle \Delta Q_{\mathbf{i}+\widehat{\alpha}/2}\right\rangle e^{i\mathbf{q}.(\mathbf{r_i}+\widehat{\alpha}/2)}, \qquad \text{(F36)}$$

$$R^{\alpha\beta}(\mathbf{q}) = \sum_{\mathbf{i}} R^{\alpha\beta}_{\mathbf{i}+\widehat{\alpha}/2-,\mathbf{j}-\widehat{\beta}/2} e^{i\mathbf{q}.(\mathbf{r_i}+\widehat{\alpha}/2-\mathbf{r_j}-\widehat{\beta}/2)}, \qquad \text{(F37)}$$

$$V_{\alpha\beta}(\mathbf{q}) = \sum_{\mathbf{i}} V_{\alpha\beta}(\mathbf{r}_{\mathbf{i}+\widehat{\alpha}/2} - \mathbf{r}_{\mathbf{j}+\widehat{\beta}/2}) e^{i\mathbf{q}.(\mathbf{r_i}+\widehat{\alpha}/2-\mathbf{r_j}-\widehat{\beta}/2)}. \qquad \text{(F38)}$$

Thus the FT $Q_\alpha(\mathbf{q})$ is defined to be bond-centered, etc.

#### 7. Simple Limits

##### *a. No LRCI*

Suppose that there is no LRCI (as in the FBMII model), then the equations become

$$Q_x(\mathbf{q}) - 2\widetilde{K}R^{xx}(\mathbf{q})Q_x(\mathbf{q}) - 2\widetilde{K}R^{xy}(\mathbf{q})Q_y(\mathbf{q}) = 0 \tag{F39}$$

$$Q_y(\mathbf{q}) + 2\widetilde{K}R^{yy}(\mathbf{q})Q_y(\mathbf{q}) + 2\widetilde{K}R^{yx}(\mathbf{q})Q_x(\mathbf{q}) = 0. \tag{F40}$$

Imagine that we are in the $q \to 0$ limit, then approximately borrowing from $q = 0$ see (F30)

$$R^{xx}(\mathbf{q}) \simeq R^{yy}(\mathbf{q}); \tag{F41}$$

$$R^{xy}(\mathbf{q}) \simeq R^{yx}(\mathbf{q}), \tag{F42}$$

so the foregoing equations become

$$Q_x(\mathbf{q}) - 2\widetilde{K}R^{xx}(\mathbf{q})Q_x(\mathbf{q}) - 2\widetilde{K}R^{xy}(\mathbf{q})Q_y(\mathbf{q}) = 0 \tag{F43}$$

$$Q_y(\mathbf{q}) - 2\widetilde{K}R^{xx}(\mathbf{q})Q_y(\mathbf{q}) - 2\widetilde{K}R^{xy}(\mathbf{q})Q_x(\mathbf{q}) = 0. \tag{F44}$$

These equations support two solutions, a monopole one

$$Q_x = Q_y, \tag{F45}$$

$$1 - 2\widetilde{K}\left(R^{xx}(\mathbf{q}) + R^{xy}(\mathbf{q})\right) = 0, \quad \text{or approximately (F31)} \tag{F46}$$

$$1 - 2\widetilde{K}\left(R^{>} + R^{<}\right) = 0, \tag{F47}$$

and a quadrupole one

$$Q_x = -Q_y, \tag{F48}$$

$$1 - 2\widetilde{K}\left(R^{xx}(\mathbf{q}) - R^{xy}(\mathbf{q})\right) = 0, \quad \text{or approximately} \tag{F49}$$

$$1 - 2\widetilde{K}\left(R^{>} - R^{<}\right) = 0. \tag{F50}$$

So in linear approximation there are two instabilities, the monopolar (F45) and quadrupolar (F48) instabilities. The monopolar instability is the strongest as it depends on the larger RF combination $(R^{>} + R^{<})$, while the quadrupolar instability depends on the weaker RF combination $(R^{>} - R^{<})$.

This result is in contrast with that in the FBMI, where the response function combinations are

$$R^{xx} \pm R^{xy} = -\sum_{\mathbf{k}} f'\left(\epsilon_{\mathbf{k}}\right) \cos\left(k_x\right) \left(\cos\left(k_x\right) \pm \cos\left(k_y\right)\right). \tag{F51}$$

Because the main weight comes from the SP's, the result (F51) is dominated by the $R_{xx} - R_{xy}$ combination, which is positive. Hence in the FBM the quadrupole solution $Q_x = -Q_y$ (F48) becomes unstable, leading to C4 symmetry breaking, first as $\widetilde{K}$ is increased (the instability in the monopole channel is much weaker in the FBMI).

The instability in the charge channel is profoundly modified by the LRCI, hence it seems that the LRCI needs to be included to make a fully physically correct extension of the FBM. This is not unexpected as the new terms in the FBMII Hamiltonian explicitly introduce charge which now must be treated properly. We shall see below that the explicit introduction of charge allows the CDW to be fully understood within the full model FBMIII.

#### b. *No Coupling to Vibrators*

Suppose we consider the opposite case $\widetilde{K} = 0$. Now the equation is

$$Q_\alpha(\mathbf{q}) = -\sum_{\beta,\gamma} R^{\alpha\gamma}(\mathbf{q}) V_{\gamma\beta}(\mathbf{q}) Q_\beta(\mathbf{q}). \tag{F52}$$

Also suppose that the bond charges can be treated as highly localized, when approximately

$$V_{\gamma\beta}(\mathbf{q}) \simeq \frac{8\pi e^2}{\epsilon v_c q^2}, \tag{F53}$$

where $v_c$ is the unit cell volume. Then defining a bond-average $Q$

$$\overline{Q}(\mathbf{q}) = \left(Q_x(\mathbf{q}) + Q_y(\mathbf{q})\right)/2, \tag{F54}$$

$$\overline{Q}(\mathbf{q}) = -\sum_{\alpha,\gamma} R^{\alpha\gamma}(\mathbf{q}) \frac{8\pi e^2}{\epsilon v_c q^2} \overline{Q}(\mathbf{q}), \quad \text{or}$$

$$1 + \frac{8\pi e^2}{\epsilon v_c q^2} \sum_{\alpha,\beta} R^{\alpha\beta}(\mathbf{q}) = 0. \tag{F55}$$

The latter equation can be simplified by taking the $q = 0$ limit of the RF, giving

$$1 + \frac{16\pi e^2}{\epsilon v_c q^2} \left(R^{>} + R^{<}\right) = 0, \tag{F56}$$

which defines the growing/decaying FT wavevector

$$q = \pm i \sqrt{\frac{16\pi e^2}{\epsilon v_c} \left(R^{>} + R^{<}\right)}. \tag{F57}$$

*c. Approximate Discussion of General Case for $q \to 0$*

We can write the generalized equations

$$Q_x(\mathbf{q}) - (2\widetilde{K}R^{xx}(\mathbf{q}) - \Pi^{xx}(\mathbf{q}))Q_x(\mathbf{q}) \\ - (2\widetilde{K}R^{xy}(\mathbf{q}) - \Pi^{xy}(\mathbf{q}))Q_y(\mathbf{q}) = 0 \tag{F58}$$

$$Q_y(\mathbf{q}) - (2\widetilde{K}R^{yy}(\mathbf{q}) - \Pi^{yy}(\mathbf{q}))Q_y(\mathbf{q}) \\ - (2\widetilde{K}R^{yx}(\mathbf{q}) - \Pi^{yx}(\mathbf{q}))Q_x(\mathbf{q}) = 0, \tag{F59}$$

where

$$\Pi^{\alpha\beta}(\mathbf{q}) = \sum_{\gamma} R^{\alpha\gamma}(\mathbf{q})V_{\gamma\beta}(\mathbf{q}). \tag{F60}$$

If we continue to assume that $q \to 0$, and for simplicity assume that the bond charges can be considered strongly localized on the $q^{-1}$ scale (a poor approximation in the 1-band model, since the quadrupolar charge ditribution is on precisely the same scale as that of the bond charges), so that the suffixes on $V_{\gamma\beta}(\mathbf{q})$ can be neglected

$$V_{\gamma\beta}(\mathbf{q}) \simeq V(\mathbf{q}), \tag{F61}$$

$$\Pi^{\alpha\beta}(\mathbf{q}) = V(\mathbf{q}) \sum_{\gamma} R^{\alpha\gamma}(\mathbf{q}), \tag{F62}$$

then the equations become (dropping the wavevector argument for clarity)

$$Q_x - (2\widetilde{K}R^{xx} - VR^{xx} - VR^{xy})Q_x \\ - (2\widetilde{K}R^{xy} - VR^{xy} - VR^{xx})Q_y = 0 \tag{F63}$$

$$Q_y - (2\widetilde{K}R^{yy} - VR^{yy} - VR^{yx})Q_y \\ - (2\widetilde{K}R^{yx} - VR^{yx} - VR^{yy})Q_x = 0. \tag{F64}$$

If we make the same $q \to 0$ approximation as before

$$R^{xx}(\mathbf{q}) \simeq R^{yy}(\mathbf{q}); \tag{F65}$$

$$R^{xy}(\mathbf{q}) \simeq R^{yx}(\mathbf{q}), \tag{F66}$$

then the foregoing equations become

$$Q_x - (2\widetilde{K}R^{xx} - VR^{xx} - VR^{xy})Q_x$$
$$- (2\widetilde{K}R^{xy} - VR^{xy} - VR^{xx})Q_y = 0 \tag{F67}$$
$$Q_y - (2\widetilde{K}R^{xx} - VR^{xx} - VR^{xy})Q_y$$
$$- (2\widetilde{K}R^{xy} - VR^{xy} - VR^{xx})Q_x = 0. \tag{F68}$$

then again the monopolar solution

$$Q_x = Q_y, \tag{F69}$$
$$1 - \Big(R^{xx}(\mathbf{q}) + R^{xy}(\mathbf{q}))(2\widetilde{K} - 2V(\mathbf{q})\Big) = 0, \tag{F70}$$
$$1 - \Big(R^{>} + R^{<})(2\widetilde{K} - 2V(\mathbf{q})\Big) = 0 \tag{F71}$$

and quadrupolar solution

$$Q_x = -Q_y, \tag{F72}$$
$$1 - 2\widetilde{K}\,(R^{>} - R^{<}) = 0. \tag{F73}$$

are supported.

The quadrupolar solution (F72) is the same as the solution without Coulomb interaction, which cancels out, it should lead to the condition for $T^*$, at least in the long-wavelength limit.

The monopolar solution can be written using the $q \to 0$ limit of the RF's (again putting $V(\mathbf{q}) = 8\pi e^2/\epsilon v_c q^2$)

$$1 - \widetilde{K}\sum_{\alpha,\beta} R^{\alpha\beta}(\mathbf{q}) + \frac{8\pi e^2}{\epsilon v_c q^2}\sum_{\alpha,\beta} R^{\alpha\beta}(\mathbf{q}) = 0, \quad \text{or} \tag{F74}$$
$$1 - 2\widetilde{K}\,(R^{>} + R^{<}) + \frac{16\pi e^2}{\epsilon v_c q^2}\,(R^{>} + R^{<}) = 0. \tag{F75}$$

In this equation if

$$2\widetilde{K}\,(R^{>} + R^{<}) > 1, \tag{F76}$$

we indeed obtain the anomalous FT where the wavevector $q$ is real, i.e. a CDW exists which does **not depend on nesting**:

$$q = \sqrt{\frac{16\pi e^2\,(R^{>} + R^{<})}{\epsilon v_c\Big(2\widetilde{K}\,(R^{>} + R^{<}) - 1\Big)}}, \tag{F77}$$

which has a large $\widetilde{K}$ limit

$$q = \sqrt{\frac{8\pi e^2}{\epsilon v_c \widetilde{K}}}. \tag{F78}$$

## 8. Summary

In the FBMI, which is missing the charge term in $Q$, having only the $X$ term, there is a $q = 0$ instability in the quadrupole, or $d$-symmetry, channel. In the FBMII, which includes the charge term in $Q$, there is a $q = 0$ instability in both the quadrupole and the monopole channels. In the FBMIII, which includes also the LRCI, there is a $q = 0$ instability in the quadrupole channel. In the monopole channel there is an anomalous Fermi-Thomas equation for the charge density or potential which describes an oscillatory, or CDW, response instead of the conventional exponentially screened response. The monopolar solution has the wavevector (F77).

There are two transition temperatures, the higher, $T_0^*$, given by the equality in (F76), defines the onset of the anomalous FT CDW-like solution. The lower temperature, $T_2^*$, defines the onset of the quadrupolar, C4 symmetry-breaking instability seen at low temperatures. This weakly wavelength-dependent instability may lock to the CDW wavelength.

---

[1] T. Timusk, Rep. Prog. Phys. **62**, 61 (1999).

[2] G. Kotliar and D. Vollhardt, Physics Today, March, 53 (2004).

[3] E. G. Maksimov , M. L. Kuli´c, and O. V. Dolgov, arXiv:0810.3789 [cond-mat].

[4] R. Daou, J. Chang, D. LeBoeuf, O. Cyr-Choinière, F. Laliberté, N. Doiron-Leyraud, B. J. Ramshaw, R. Liang, D. A. Bonn, W. N. Hardy, and L. Taillefer, Nature **463**, 519 (2010).

[5] Y. Kohsaka, C. Taylor, P. Wahl, A. Schmidt, J. Lee, K. Fujita, J. W. Alldredge, K. McElroy, J. Lee, H. Eisaki, S. Uchida, D.-H. Lee, and J. C. Davis, Nature **454**, 1072 (2008).

[6] Y. Kohsaka, C. Taylor, K. Fujita, A. Schmidt, C. Lupien, T. Hanaguri, M. Azuma, M. Takano, H. Eisaki, H. Takagi, S. Uchida, and J. C. Davis, Science **315**, 1380 (2007).

[7] K. McElroy K, J. Lee, J. A. Slezak, D.-H. Lee, H. Eisaki H, S. Uchida, and J. C. Davis, Science **309**, 1048 (2005).

[8] D. M. Newns and C. C. Tsuei, Nature Physics **3**, 184 (2007).

[9] T. Sakai, D. Poilblanc, and D. J. Scalapino, Phys. Rev. B. **55**, 8445 (1997).

[10] C. C. Tsuei and J. R. Kirtley, Rev. Mod. Phys. **72**, 969 (2000).

[11] A. Bussman-Holder and A. R. Bishop, Phys. Rev. B **44**, 2853 (1991).

[12] V. H. Crespi and M. L. Cohen Phys. Rev. B **48**, 398 (1993).

[13] G. D. Mahan, Phys. Rev. B **56**, 8322 (1997).

[14] D. J. Pringle, G. V. M. Williams, and J. L. Tallon, Phys. Rev. B **62**, 12527 (2000).

[15] D. Zech, H. Keller, K. Conder, E. Kaldis, E. Liarokapis, N. Poulakis, and K. A. Müller, Nature **371**, 681 (1994).

[16] K. C. Hewitt, X. K. Chen, C. Roch, J. Chrzanowski, J. C. Irwin, E. H. Altendorf, R. Liang, D. Bonn, and W. N. Hardy, Phys. Rev. B **69**, 064514 (2004).

[17] L. Pintschovius, Phys. Stat. Sol. (b) **242**, 30 (2005).

[18] F. Giustino, M. L. Cohen, and S. G. Louie, Nature **452**, 975 (2008).

[19] R. Heid, K.-P. Bohnen, R. Zeyher, and D. Manske, Phys. Rev. Lett. **100**, 137001 (2008).

[20] R. Car and M. Parrinello, Phys. Rev. Lett. **55**, 2471 (1985).

[21] A. Bianconi, N. L. Saini, A. Lanzara, M. Missori, T. Rossetti, H. Oyanagi, H. Yamaguchi, K. Oka, and T. Ito, Phys. Rev. Lett. **76**, 3412 (1996).

[22] W. E. Pickett, R. E. Cohen, and H. Krakauer, Phys. Rev. Lett. **67**, 228 (1991).

[23] R. A. Nistor, Ph. D. Thesis, University of Western Ontario, p. 114-116 (2009).

[24] E. Pavarini, I. Dasgupta, T. Saha-Dasgupta, O. Jepsen, and O. K. Andersen, Phys. Rev. Lett. **87**, 047003 (2001).

[25] R. S. Markiewicz, J. Phys. Chem. Solids **58**, 1179 (1997).

[26] A. A. Abrikosov, K. Gofron, J. C. Campuzano, M. Lindroos, A. Bansil, H. Ding, D. D. Koelling, and B. Dabrowski, Phys. Rev. Lett. **73**, 3302 (1994).

[27] M. Le Tacon, A. Sacuto, A. Georges, G. Kotliar, Y. Gallais, D. Colson, and A. Forget, Nature Physics **2**, 537 (2006).

[28] A. G. Loeser, Z.-X. Shen, D. S. Dessau, D. S. Marshall, C. H. Park, P. Fournier, and A. Kapitulnik, Science **273**, 325 (1996).

[29] H. Ding, T. Yokoya, J. C. Campuzano, T. Takahashi, M. Randeria, M. R. Norman, T. Mochikupara, K. Kadowaki, and J. Giapintzakis Nature **382**, 51 (1996).

[30] J. Lee, K. Fujita, K. McElroy, J. A. Slezak, M. Wang, Y. Aiura, H. Bando, M. Ishikado, T. Masui, J.-X. Zhu, A. V. Balatsky, H. Eisaki, S. Uchida, and J. C. Davis, Nature **442**, 546

(2006).

[31] J. A. Robertson, S. A. Kivelson, E. Fradkin, A. C. Fang, and A. Kapitulnik, Phys. Rev. B **74**, 134507 (2006).

[32] M. R. Norman, A. Kanigel, M. Randeria, U. Chatterjee, and J. C. Campuzano JC, Phys. Rev. B **76**, 174501 (2007); M. Norman, Nature **392**, 157 (1998).

[33] A. Kanigel, M. R. Norman, M. Randeria, U. Chatterjee, S. Souma, A. Kaminski, H. M. Fretwell, S. Rosenkranz, M. Shi, T. Sato, T. Takahashi, Z. Z. Li, H. Raffy, K. Kadowaki, D. Hinks, L. Ozyuzer, and J. C. Campuzano, Nature Physics **2**, 447 (2006).

[34] X. J. Zhou *et al.* Phys. Rev. Lett. **92**, 187001 (2004).